%% file: main.tex
\documentclass[sigconf]{acmart}
\AtBeginDocument{%
  }

\setcopyright{acmlicensed}
\copyrightyear{2026}
\acmYear{2026}
\acmDOI{XXXXXXX.XXXXXXX}
\acmISBN{978-1-4503-XXXX-X/2026/11}
\usepackage{booktabs}
\usepackage{multirow}
\usepackage{array}
\usepackage{tabularx}
\usepackage{mathtools}
\usepackage{xcolor}
\usepackage{colortbl}
\usepackage{placeins}
\usepackage{stfloats}
\usepackage{enumitem}
\usepackage{algorithm}
\usepackage{algpseudocode}
\usepackage{makecell}
\renewcommand\footnotetextcopyrightpermission[1]{}

\setlist[itemize]{leftmargin=1.2em}
\renewcommand{\topfraction}{0.95}
\renewcommand{\bottomfraction}{0.95}
\renewcommand{\textfraction}{0.05}
\renewcommand{\floatpagefraction}{0.85}
\setcitestyle{numbers,sort,nocompress,square,comma}
\newcommand{\method}{STAP}

\newcommand{\R}{\mathbb{R}}
\newcommand{\promptbox}[1]{%
  \noindent\fcolorbox{black!12}{gray!6}{%
  \parbox{\dimexpr\linewidth-2\fboxsep-2\fboxrule\relax}{#1}}%
}

\begin{document}

\title{Seeing Further and Wider: Joint Spatio-Temporal Enlargement for Micro-Video Popularity Prediction}

\author{Dali Wang}
\authornote{Equal contribution.}
\affiliation{%
  \institution{Huazhong University of Science and Technology}
  \city{Wuhan}
\country{China}}

\author{Yunyao Zhang}
\authornotemark[1]
\affiliation{%
  \institution{Huazhong University of Science and Technology}
  \city{Wuhan}
\country{China}}

\author{Junqing Yu}
\affiliation{%
  \institution{Huazhong University of Science and Technology}
  \city{Wuhan}
\country{China}}

\author{Yi-Ping Phoebe Chen}
\affiliation{%
  \institution{La Trobe University}
  \city{Melbourne}
\country{Australia}}

\author{Chen Xu}
\affiliation{%
  \institution{Beijing Institute of Computer Technology and Applications}
  \city{Beijing}
  \country{China}
}

\author{Zikai Song}
\authornote{Corresponding author: \nolinkurl{skyesong@hust.edu.cn}.}
\affiliation{%
  \institution{Huazhong University of Science and Technology}
  \city{Wuhan}
\country{China}}

\renewcommand{\shortauthors}{Wang et al.}

\makeatletter
\fancypagestyle{standardpagestyleclean}[standardpagestyle]{%
  \fancyhead[LE,RO]{}%
}
\makeatother

\begin{abstract}

  Micro-video popularity prediction (MVPP) aims to forecast the future popularity of videos on online media, which is essential for applications such as content recommendation and traffic allocation.
  In real-world scenarios, it is critical for MVPP approaches to understand both the temporal dynamics of a given video (temporal) and its historical relevance to other videos (spatial).
  However, existing approaches suffer from limitations in both dimensions: temporally, they rely on sparse short-range sampling that restricts content perception; spatially, they depend on flat retrieval memory with limited capacity and low efficiency, hindering scalable knowledge utilization.
  To overcome these limitations, we propose a \textbf{unified framework that achieves joint spatio-temporal enlargement}, enabling precise perception of extremely long video sequences (e.g., up to 100 frames, far beyond the typical 20-frame window) while supporting a scalable memory bank that can infinitely expand to incorporate all relevant historical videos.
  Technically, we employ a \textbf{Temporal Enlargement} driven by a frame scoring module that extracts highlight cues from video frames through two complementary pathways: sparse sampling and dense perception. Their outputs are adaptively fused to enable robust long-sequence content understanding. For \textbf{Spatial Enlargement}, we construct a Topology-Aware Memory Bank that hierarchically clusters historically relevant content based on topological relationships. Instead of directly expanding memory capacity, we update the encoder features of the corresponding clusters when incorporating new videos, enabling unbounded historical association without unbounded storage growth.
  Extensive experiments on three widely used MVPP benchmarks demonstrate that our method consistently outperforms 11 strong baselines across mainstream metrics, achieving robust improvements in both prediction accuracy and ranking consistency.

\end{abstract}

\vspace{-10pt}

\maketitle
\pagestyle{standardpagestyleclean}


\section{Introduction}
\begin{figure}[t]
  \centering
  \includegraphics[width=0.98\linewidth]{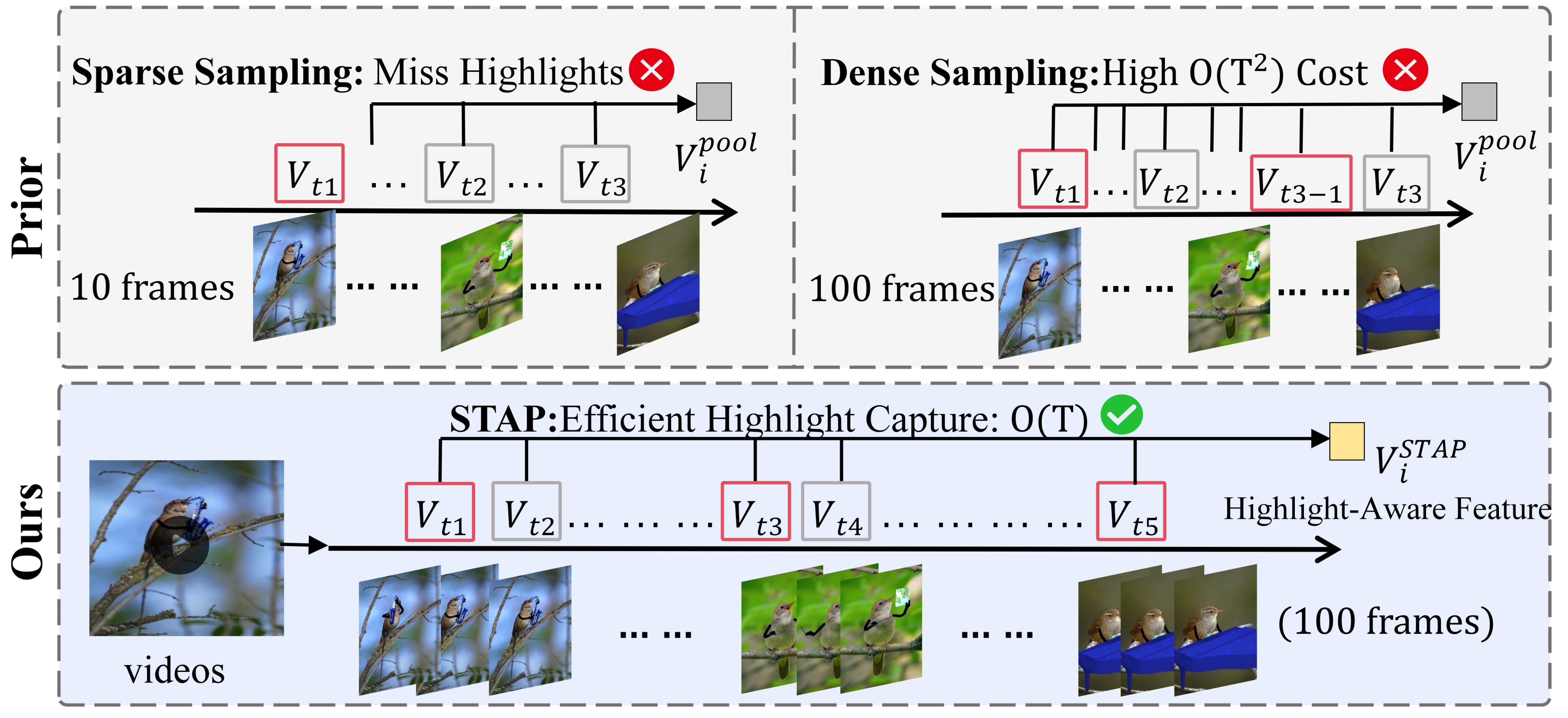}
  \vspace{2pt}
  \makebox[\linewidth][c]{\small (a) Innovation I: Temporal Enlargement with efficient highlight capture.}
  \includegraphics[width=0.98\linewidth]{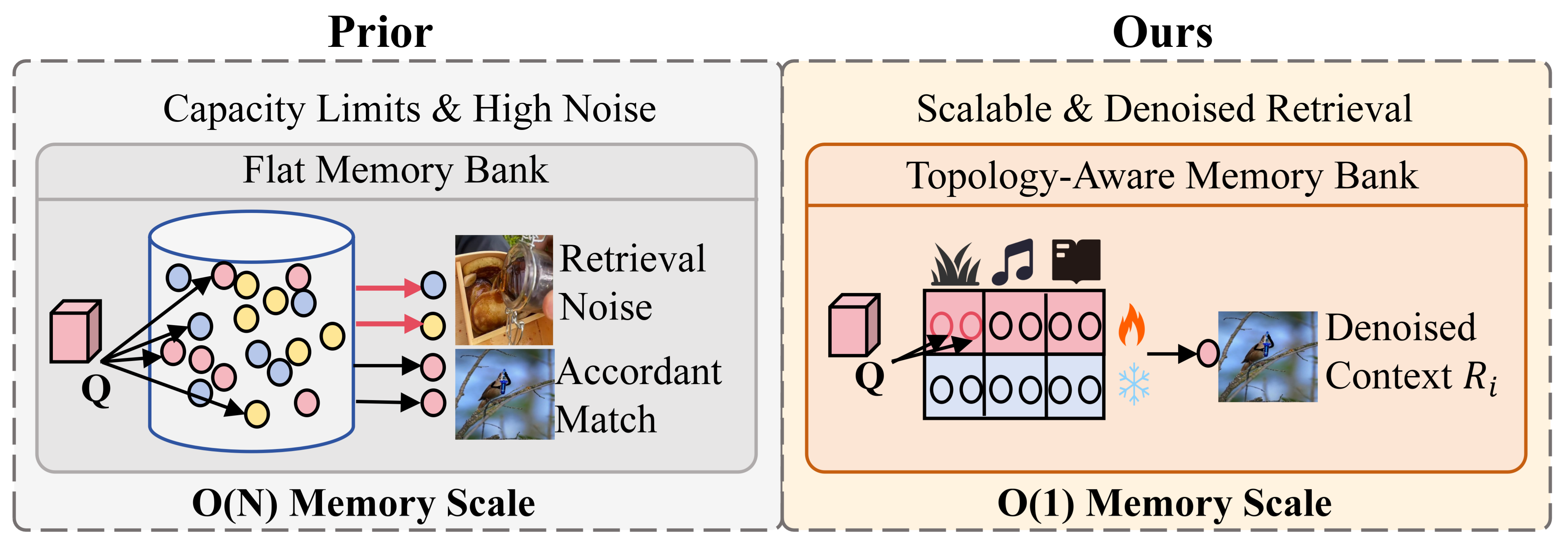}
  \makebox[\linewidth][c]{\small (b) Innovation II: Topology-Aware Memory for scalable denoised retrieval.}
  \caption{Motivation of our work. (a) Innovation I addresses temporal modeling by preserving long-range highlights with linear-complexity processing. (b) Innovation II addresses spatial modeling by using a Topology-Aware Memory Bank for denoised and scalable retrieval.}
  \Description{Stacked two-part motivation figure. Part (a) presents Innovation I on temporal enlargement, contrasting prior sparse or dense sampling with our efficient highlight capture. Part (b) presents Innovation II on spatial enlargement, contrasting flat memory retrieval with topology-aware denoised retrieval at constant memory scale.}
  \label{fig:motivation}
\end{figure}

Micro-video content has become a major modality on contemporary social platforms. Micro-video Popularity Prediction (MVPP) aims to forecast the future popularity of such content by jointly modeling its multimodal signals (e.g., visual, audio, and textual information)~\cite{chen2016micro,wu2016temporal,cheung2022crossmodal}, making it a central problem in multimedia intelligence.
Beyond its direct utility for recommendation, traffic allocation, and risk-aware content governance, MVPP also serves as a representative benchmark for long-context multimodal modeling, cross-modal alignment, and retrieval-augmented prediction.

Recent representative methods in MVPP typically adopt a \textbf{two-stage paradigm}~\cite{xu2020attention,mao2023catboost,zhong2024mmra,cheng2025scrag}. The \textbf{\textit{first stage}} focuses on multimodal fusion~\cite{xu2020attention,mao2023catboost,ye2025mvp}, which jointly models visual, audio, and textual modalities to capture the characteristics of the current video. Common practices include designing cross-modal attention mechanisms~\cite{xu2020attention,ye2025mvp} or hierarchical fusion networks~\cite{mao2023catboost, song13} to extract more discriminative representations. The \textbf{\textit{second stage}} introduces retrieval augmentation~\cite{zhong2024mmra,cheng2024icpf,cheng2025scrag}, where a knowledge base is constructed from historical samples. During prediction, the most relevant historical samples are retrieved and their features are incorporated as external knowledge into the prediction module~\cite{zhong2024mmra,cheng2024rah,cheng2024icpf,cheng2025scrag}, enabling the model to better adapt to real-world deployment. Together, these two modules enable the model to accurately perceive the current video’s content while effectively leveraging historical associations, jointly improving prediction robustness and generalization.

Existing MVPP approaches face \textbf{two fundamental challenges}:
(\textbf{C1}) \textbf{\textit{Temporally}}, long-duration micro-videos contain key semantic highlights that are sparsely distributed across the full timeline. Existing methods typically model only a small set of sparsely sampled frames, relying on fixed sampling and short-window modeling. This inevitably discards critical long-range cues, causing significant degradation in content perception for longer videos;
(\textbf{C2}) \textbf{\textit{Spatially}}, historical relevance provides valuable context for popularity prediction. Existing retrieval-augmented methods construct flat memory banks that scale linearly with the number of training samples. This design suffers from rapidly rising matching costs, increased retrieval noise, and weak generalization on cold-start subsets as memory scales.

\textbf{\textit{These two challenges are intrinsically coupled in a negative feedback loop}}: poor temporal perception degrades the quality of query representations for retrieval, while noisy and inefficient spatial retrieval introduces misleading context into temporal modeling. Consequently, optimizing either dimension limited gains and remains insufficient for real-world MVPP deployment.

To address these challenges, we propose \textbf{STAP} (\textbf{S}patial-\textbf{T}emporal \textbf{A}ugmentation for \textbf{P}opularity Prediction), a unified framework built on the core principle of joint spatio-temporal enlargement. For \textbf{Temporal Enlargement}(for \textbf{C1}), we employ a frame scoring module that extracts highlight cues through two complementary pathways: sparse sampling and dense perception. Their outputs are adaptively fused to enable robust long-sequence content understanding, capturing up to 100 frames, which is far beyond the typical 20-frame window. For \textbf{Spatial Enlargement}(for \textbf{C2}), we construct a Topology-Aware Memory Bank that hierarchically clusters historically relevant content based on topological relationships. Instead of directly expanding memory capacity, we update the encoder features of the corresponding clusters when incorporating new videos, enabling unbounded historical association without unbounded storage growth. The two modules are jointly optimized in an end-to-end pipeline to break the negative feedback loop between temporal perception and scalable retrieval. Extensive experiments on three widely used datasets (MicroLens, SMPD-video, Informs) demonstrate that STAP consistently outperforms competitive baselines across mainstream evaluation metrics.

\enlargethispage{2\baselineskip}
The main contributions are summarized as follows:
\begin{itemize}[topsep=0pt,itemsep=1pt,parsep=0pt,partopsep=0pt]
  \item \textbf{Unified Spatio-Temporal Framework.} We propose STAP, the first MVPP framework that jointly optimizes temporal and spatial enlargement in an end-to-end manner. By integrating long-context temporal perception with scalable structured retrieval, it breaks the negative coupling that limits existing methods.
  \item \textbf{Temporal Enlargement Module.} We design a temporal module that extracts highlight cues through two complementary pathways (sparse sampling and dense perception), and adaptively fuses their outputs. This enables robust long-sequence modeling and effectively captures sparsely distributed highlights in long micro-videos.
  \item \textbf{Spatial Enlargement Module.} We construct a Topology-Aware Memory Bank that hierarchically clusters historically relevant content. Instead of expanding memory capacity linearly, we update the encoder features of the corresponding clusters when incorporating new videos, enabling unbounded historical association without unbounded storage growth.
\end{itemize}

\begin{figure*}[!t]
  \centering
  \includegraphics[width=0.92\textwidth]{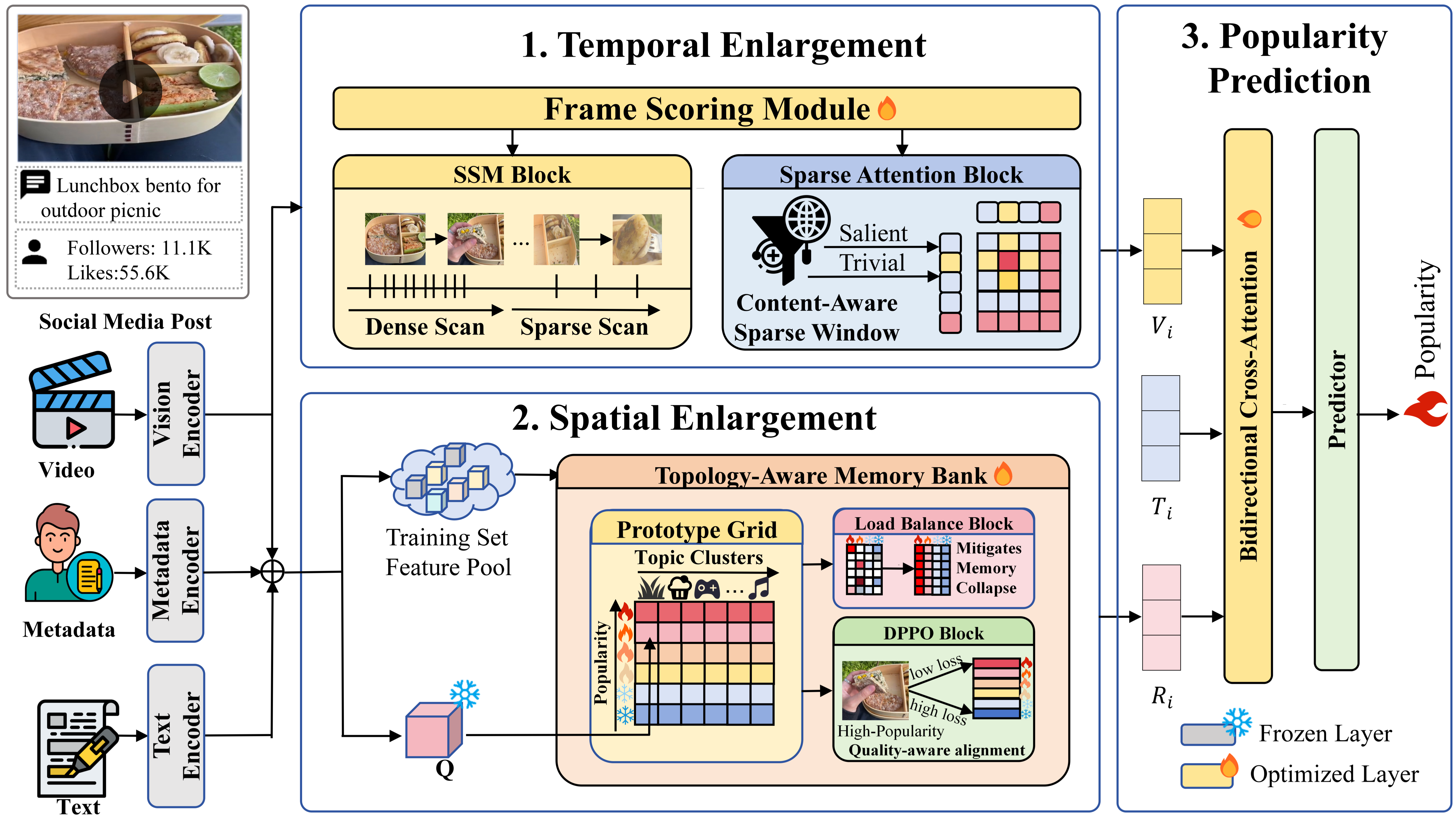}
  \caption{Overall framework of \method{}. (1) The Temporal Enlargement process starts from multimodal encoding and frame scoring, then performs dual-path temporal modeling where the SSM branch uses dynamic dense/sparse scan to preserve temporal order and the Sparse Attention branch captures salient interactions in content-aware sparse windows, producing visual feature $V_i$. (2) The Spatial Enlargement process constructs and queries a Topology-Aware Memory Bank from the training feature pool, retrieves top-$K$ prototypes through Prototype-Grid routing, and refines routing with Load Balance and DPPO blocks to generate retrieval feature $R_i$. (3) The Popularity Prediction process fuses $V_i$, $T_i$, and $R_i$ with bi-directional cross-attention and feeds the fused representation into the predictor to estimate final popularity.}
  \Description{Detailed workflow of STAP: multimodal encoding and fusion, temporal dual-path modeling, topology-aware memory retrieval, and final popularity prediction.}
  \label{fig:overview}
  \vspace{-4pt}
\end{figure*}

\section{Related Work}

\noindent\textbf{Micro-video Popularity Prediction.}
MVPP methods mainly follow two paradigms. Feature-engineering methods manually construct modality-specific cues from visual content, metadata, and engagement statistics, then apply classical predictors for popularity estimation~\cite{khosla2014makes,wu2016temporal,lai2020hyfea,hsu2023mftm}. Representative methods include temporal-dynamics decomposition for popularity evolution~\cite{wu2016temporal}, top-performing challenge systems based on feature aggregation and ensemble design~\cite{lai2020hyfea}, feature-intensive boosting pipelines that improve robustness on SMP benchmarks~\cite{hsu2023mftm}, and diffusion-inspired popularity propagation modeling~\cite{cheng2024pdiff}. These methods are interpretable but rely heavily on manual feature construction and often struggle with complex nonlinear cross-modal interactions.
Deep multimodal methods learn end-to-end representations and generally provide stronger cross-modal integration, including transductive multimodal learning~\cite{chen2016micro}, attention-based interaction~\cite{xu2020attention,cheung2022crossmodal}, and variational modeling for latent uncertainty control (e.g., MASSL and HMMVED)~\cite{zhang2022massl,xie2023hmmved}. Recent multimodal fusion studies further show that stronger cross-modal coupling and alignment can substantially improve representation quality, as evidenced by coupled state-space fusion and language-guided visual alignment designs~\cite{li2024coupled,song2024autogenic}. More recent retrieval-augmented methods, such as MMRA, ICPF, and SCRAG, inject historical context or retrieval-guided completion signals to improve prediction quality under challenging settings~\cite{zhong2024mmra,cheng2024icpf,cheng2025scrag}. In-context prompting lines in MVPP are also related to broader few-shot and prompt-learning literature across language and vision-language models~\cite{brown2020language,liu2023pre,zhou2022learning,jia2022visual,khattak2023maple,vaswani2017attention,devlin2019bert}. Meanwhile, adjacent work on social-network simulation and reasoning suggests that future popularity modeling may also benefit from richer context modeling beyond video content alone~\cite{zhang-etal-2025-ga,zhang2026couplingmacrodynamicsmicro,zhang2026intervensiminterventionawaresocialnetwork,zhang2026semanticawarelogicalreasoningsemiotic,zhang2026logicalphasetransitionsunderstanding}. Despite these advances, current approaches still show limited long-range temporal perception and weak retrieval scalability under large, noisy MVPP settings; \method{} addresses this gap with joint Temporal Enlargement and Spatial Enlargement.

\noindent\textbf{Long-Context Retrieval Modeling.}
Recent MVPP research combines long-context temporal understanding with retrieval-based context enhancement~\cite{li2025taco, Retrack, HABIT, OFFSET, INTENT}. On the temporal side, Longformer introduces sparse attention patterns to reduce quadratic complexity, while S4/Mamba-family state-space modeling offers strong long-range dependency capture with near-linear computation; VideoMamba further validates this design in video scenarios~\cite{beltagy2020longformer,gu2021s4,gu2023mamba,lin2024videomamba}. Related long-range video backbones and efficient-attention variants include I3D~\cite{carreira2017i3d}, SlowFast~\cite{feichtenhofer2019slowfast}, TSM~\cite{lin2019tsm}, TimeSformer~\cite{bertasius2021timesformer}, ViViT~\cite{arnab2021vivit}, BigBird~\cite{zaheer2020bigbird}, Linformer~\cite{wang2020linformer}, Performer~\cite{choromanski2021performer}, and Nystr\"omformer~\cite{xiong2021nystromformer}. Recent vision studies further indicate that cyclic-window attention, compact masked transformer design, and temporally coherent flow modeling are effective for preserving salient structure under efficiency constraints~\cite{song2022transformer,song2023compact,song2025temporal, song15, song11, li2025miv}. Self-supervised fine-tuning of Video-LLMs also highlights the value of fragment-level training for fine-grained long-video understanding~\cite{hu2025sf2t}. On the retrieval side, MMRA retrieves semantically related historical samples~\cite{zhong2024mmra}; ICPF leverages retrieved examples as in-context prompts~\cite{cheng2024icpf}; and RAH/SCRAG enrich prediction via retrieval-enhanced structure modeling or retrieval-guided completion~\cite{cheng2024rah,cheng2025scrag}, which is conceptually consistent with retrieval-augmented generation and dense/late-interaction retrieval paradigms~\cite{lewis2020rag,karpukhin2020dpr,khattab2020colbert, MELT, STABLE, HINT, REFINE}. Viewing model parameters as implicit key-value memory also motivates our parameterized memory design for controllable retrieval~\cite{geva2021transformer}.\\
\noindent However, existing pipelines still couple these components loosely: temporal encoders are rarely optimized for MVPP's sparse-highlight dynamics, and retrieval modules often depend on flat memory structures that incur rising cost, stronger noise interference, and unstable routing under scale~\cite{guo2020avq}. Our \method{} treats this challenge as a unified objective by jointly optimizing Temporal Enlargement and topology-aware Spatial Enlargement in one end-to-end framework.

\section{Methodology}
In this section, we first formalize micro-video popularity prediction with a unified notation system, and then present the design of \method{} with emphasis on joint spatio-temporal enlargement.
\textbf{Problem Definition} formalizes the learning objective, \textbf{Temporal Enlargement} captures long-range highlight dynamics, \textbf{Spatial Enlargement} provides scalable topology-aware retrieval context, and \textbf{Popularity Prediction Procedure} fuses multimodal and retrieved features for final prediction.
Figure~\ref{fig:overview} illustrates the key components of \method{} and their relationships in a unified pipeline.
\subsection{Problem Definition}
\noindent\textbf{Innovation Rationale.}
Before defining the task mathematically, we briefly justify the theoretical strengths of our two innovations in the \method{} setting for popularity prediction. The core challenge is not only multimodal fusion, but also how to maintain effective information flow under long temporal context, noisy supervision, and long-tail popularity distributions.

\noindent\textbf{Temporal Advantage.}
Temporal Enlargement is designed for the sparse-highlight nature of micro-videos, where only a small portion of frames contributes strongly to final popularity while many frames are redundant. Uniform sampling or fixed-step modeling therefore creates an inherent bias: either important late-stage cues are missed, or computational budget is wasted on low-value segments. Our Frame Scoring Module with anchor guidance plus dual temporal blocks addresses this mismatch. The SSM Block preserves long-range temporal order through dynamic $\Delta$ dense/sparse scanning, while the Sparse Attention Block explicitly builds non-local links among distant highlight frames through content-aware sparse windows. This complementary design improves signal retention and noise suppression simultaneously, offering a stronger quality--efficiency trade-off than short-window temporal modeling.

\noindent\textbf{Spatial Advantage.}
Spatial Enlargement targets a second structural bottleneck: retrieval instability at scale. In flat retrieval memory, matching quality degrades as memory grows because head samples dominate while semantically relevant tail samples are underrepresented. Our Topology-Aware Memory Bank organizes retrieval space by popularity partitions and topic clusters in a Prototype-Grid, introducing a more suitable inductive bias for popularity prediction: semantic similarity and popularity prior are modeled in a structured way rather than mixed in one global pool. Sparse top-$K$ routing controls complexity, the Load Balance Block reduces memory-slot collapse, and the DPPO Block aligns routing preference with pairwise ranking supervision. Together, these mechanisms improve retrieval discrimination, ranking consistency, and scalability under long-tail data.

\noindent\textbf{Task Formulation.}
Let $\mathcal{V}=\{\mathcal{S}_1,\ldots,\mathcal{S}_N\}$ denote a micro-video corpus with $N$ samples, where each sample is represented as $\mathcal{S}_i=(V_i,T_i,U_i,Y_i)$: $V_i$ is the temporal visual representation produced by Temporal Enlargement from raw frame sequence $X_i$, $T_i$ is the associated text metadata, $U_i$ denotes metadata signals, and $Y_i \in \R$ is the target popularity value over a predefined future horizon. Specifically, for MicroLens, \method{} predicts cumulative views $Y_i$ of a given video; for SMPD-video, $Y_i$ follows the official challenge-provided popularity score; for Informs, $Y_i$ is a normalized composite value of likes, comments, plays, and shares. The goal of micro-video popularity prediction is to learn a function that predicts $\hat{Y}_i$ from multimodal inputs.

Spatial Enlargement outputs a retrieval-augmented feature vector after sparse top-$K$ routing:
\begin{equation}
  \mathbf{R}_i = \mathcal{R}_{\psi}(V_i,T_i,U_i;\mathcal{M}) \in \R^{d_R},
\end{equation}
where $\mathcal{M}=\{m_{p,c}\} \in \R^{P\times C\times d_m}$ is a Topology-Aware Memory Bank and $\mathcal{R}_{\psi}(\cdot)$ is the Spatial Enlargement retrieval operator. Specifically, $p$ indexes popularity-level partitions and $c$ indexes topic clusters within each partition. The popularity prediction is
\begin{equation}
  \hat{Y}_i = f_{\phi}\!\left(V_i,T_i,U_i,\mathbf{R}_i\right).
\end{equation}

Given training set $\mathcal{D}=\{(V_i,T_i,U_i,Y_i)\}_{i=1}^{N}$, the learning objective is
\begin{equation}
  \min_{\phi,\psi}\;\frac{1}{N}\sum_{i=1}^{N}\mathcal{L}\big(f_{\phi}(V_i,T_i,U_i,\mathbf{R}_i),Y_i\big),
\end{equation}
where $\mathcal{L}(\cdot)$ is a multi-term training objective and $\mathbf{R}_i$ is produced by $\mathcal{R}_{\psi}$.
\subsection{Temporal Enlargement}
A unique characteristic of micro-videos is that popularity-dominant semantic highlights are sparsely distributed across the timeline, with highly uneven semantic density across frames. Fixed sparse sampling and short-window modeling miss these critical cues, while dense full-sequence modeling brings prohibitive computational cost. To address this, we design a block-aligned temporal pipeline with four components: Frame Scoring Module, SSM Block, Sparse Attention Block, and Gated Fusion Mechanism.

\begin{figure}[t]
  \centering
  \includegraphics[width=0.95\linewidth]{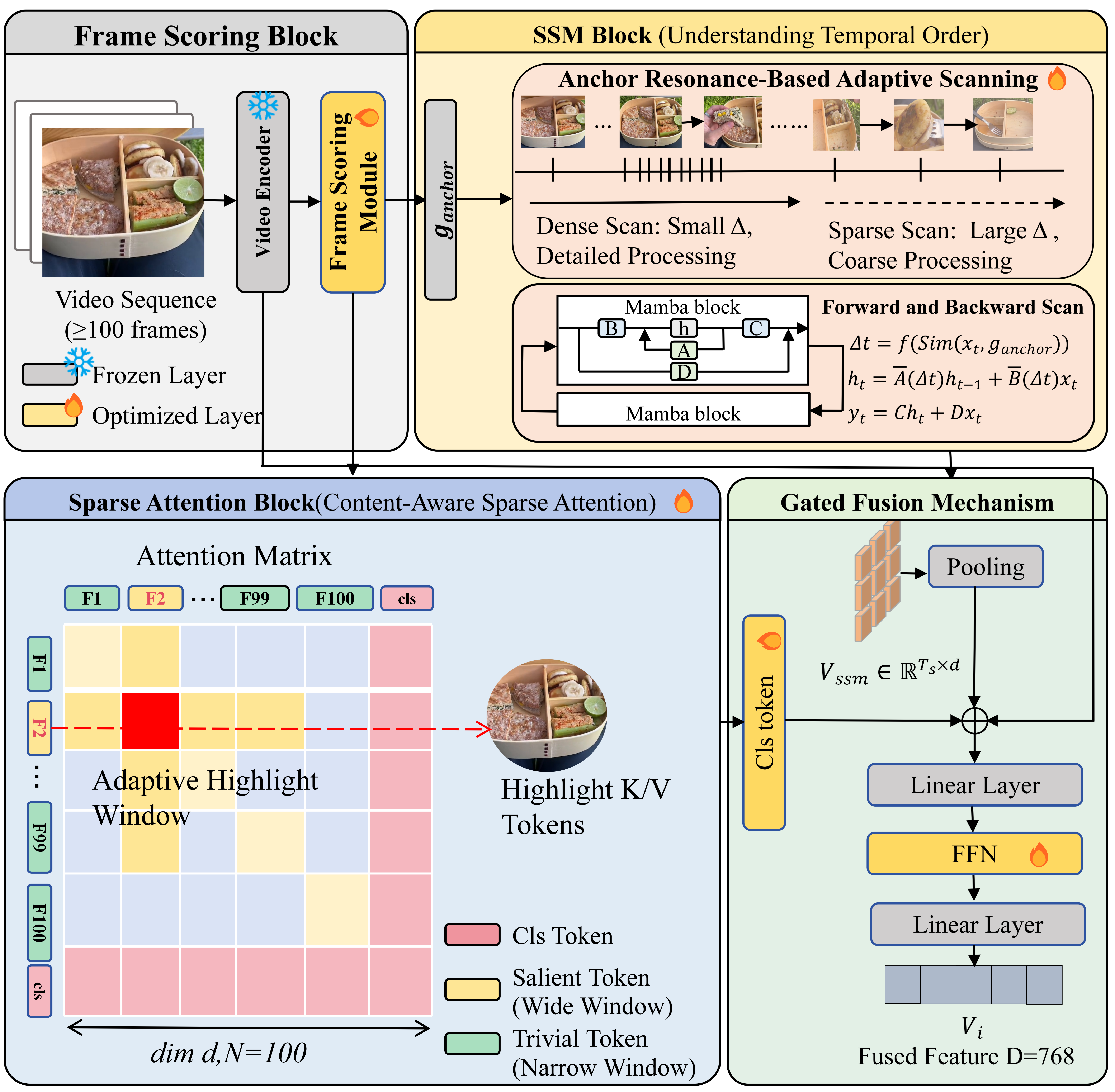}
  \caption{Temporal Enlargement workflow. Frame scoring first builds importance-aware guidance and a global anchor; then the SSM branch performs dynamic dense/sparse scan and the Sparse Attention branch models salient interactions in adaptive windows; finally, gated fusion combines both outputs into temporal representation $V_i$.}
  \Description{Temporal workflow from frame scoring to dual-path encoding and gated fusion.}
  \label{fig:temporal_pipeline}
  \vspace{-4pt}
\end{figure}

\noindent\textbf{Frame Scoring Module.}
This module is designed to estimate frame-level salience and produce a shared global anchor that guides both temporal pathways.
As illustrated in Figure~\ref{fig:temporal_pipeline}, the Frame Scoring Module is a lightweight two-layer MLP. For frame $t$, we first compute inter-frame variation $\Delta x_t=x_t-x_{t-1}$ (with $\Delta x_1=\mathbf{0}$), and then obtain the pre-activation score by $u_t = W_2\,\mathrm{GELU}(W_1\Delta x_t+b_1)+b_2$.
The normalized importance weight is computed by $w_t=\sigma(u_t)$ with $w_t\in(0,1)$.
Frames with larger variation receive higher weights, while redundant static frames are down-weighted. The shared global semantic anchor is
{\setlength{\abovedisplayskip}{5pt}
  \setlength{\belowdisplayskip}{5pt}
  \setlength{\abovedisplayshortskip}{5pt}
  \setlength{\belowdisplayshortskip}{5pt}
  \begin{equation}
    g_{\text{anchor}}=\tanh\!\left(\mathrm{LN}\!\left(\frac{\sum_{t=1}^{T} w_t x_t}{\sum_{t=1}^{T} w_t+\epsilon}\right)\right).
  \end{equation}
}
This anchor coordinates both temporal pathways with a consistent global signal.

\noindent\textbf{SSM Block.}
This block is designed to preserve long-range temporal order with adaptive dense/sparse scanning, serving as the temporal backbone of \method{}.
We model anchor resonance-based adaptive scanning with a selective SSM:
{\setlength{\abovedisplayskip}{5pt}
  \setlength{\belowdisplayskip}{5pt}
  \setlength{\abovedisplayshortskip}{5pt}
  \setlength{\belowdisplayshortskip}{5pt}
  \begin{equation}
    h_t=\bar{A}(\Delta_t)h_{t-1}+\bar{B}(\Delta_t)x_t,\quad
    y_t=Ch_t+Dx_t.
  \end{equation}
}
The dynamic step size is controlled by anchor--frame similarity and frame importance, i.e., $\Delta_t=\Delta_{\text{base}}+\alpha\,\mathrm{Sim}(g_{\text{anchor}},x_t)+\rho\,w_t$.
Small $\Delta_t$ triggers dense scan (detailed processing) on salient frames, while large $\Delta_t$ triggers sparse scan (coarse processing) on trivial spans. We further adopt bidirectional SSM scan and fuse forward/backward outputs as $y_t^{\text{ssm}}=y_t^{\rightarrow}+y_t^{\leftarrow}$. With selective-scan structured parameterization, this block has time complexity $\mathcal{O}(Td_h)$ and memory complexity $\mathcal{O}(Td_h)$.

\noindent\textbf{Sparse Attention Block.}
This block is designed to capture non-local highlight interactions efficiently, complementing the order-focused SSM dynamics.
To capture long-range highlight interactions with sub-quadratic cost, we build a content-aware sparse-attention block with an adaptive highlight window. For each frame, we first obtain query/key/value projections, and then restrict interaction to an adaptive neighborhood $\mathcal{N}(t)=\{j\,|\,|j-t|\leq \mathrm{win}_t\}$, where $\mathrm{win}_t=w_{\text{base}}+\beta\lVert x_t\rVert_2$. Salient tokens are assigned wider windows and contribute more highlight K/V tokens, while trivial tokens keep narrow windows. Attention is computed only inside $\mathcal{N}(t)$:
\begin{equation}
  y_t^{\text{attn}}=
  \sum_{j \in \mathcal{N}(t)}
  \frac{\exp(q_t^\top k_j/\sqrt{d_a})}{\sum_{m \in \mathcal{N}(t)}\exp(q_t^\top k_m/\sqrt{d_a})}v_j.
\end{equation}
If $\bar{w}$ is the average window size, complexity is $\mathcal{O}(T\bar{w}d_a)$ versus $\mathcal{O}(T^2d_a)$ for dense attention.

\noindent\textbf{Gated Fusion Mechanism.}
This mechanism is designed to adaptively integrate complementary temporal cues and residual visual evidence into a robust final representation.
We keep both pathways decoupled before fusion, then obtain temporal representation by gated residual integration:
\begin{equation}
  V_i=\gamma_sY_i^{\text{ssm}}+\gamma_aY_i^{\text{attn}}+\gamma_g g_{\text{anchor}}+X_i^{\text{proj}},
\end{equation}
where $X_i^{\text{proj}}=\mathrm{Pool}(W_{\text{res}}X_i+b_{\text{res}})$ is the residual projection of the original frame sequence, preserving low-level visual cues that may be attenuated by temporal transformations. $(\gamma_s,\gamma_a,\gamma_g)$ are dynamic fusion weights produced by the gating module. All temporal parameters are optimized end-to-end by backpropagation from the final popularity objective.

In particular, the Frame Scoring Module parameters $(W_1,W_2,b_1,b_2)$, the SSM Block parameters (including state transition/projection parameters and step-size modulation coefficients), and the Sparse Attention Block parameters (query/key/value projections and adaptive-window-related factors) are all differentiable and trainable. Their gradients are propagated from the same final prediction loss, so these temporal components are fine-tuned jointly with memory slots in Spatial Enlargement under a unified end-to-end optimization process.
\subsection{Spatial Enlargement}
In real-world popularity prediction, data typically follow a pronounced long-tail distribution: head samples dominate user interactions, while tail samples remain sparse and under-represented. Flat retrieval memory with linear memory growth therefore amplifies retrieval noise, increases matching cost, and weakens generalization on cold-start subsets. To address this, we design a block-aligned spatial pipeline with four components: Offline Memory Bank Initialization Pipeline, Prototype-Grid Routing and Top-$K$ Retrieval, Load Balance Block, and DPPO Block, enabling scalable knowledge utilization by updating matched cluster features instead of linearly appending memory slots.

\begin{figure}[t]
  \centering
  \includegraphics[width=0.95\linewidth]{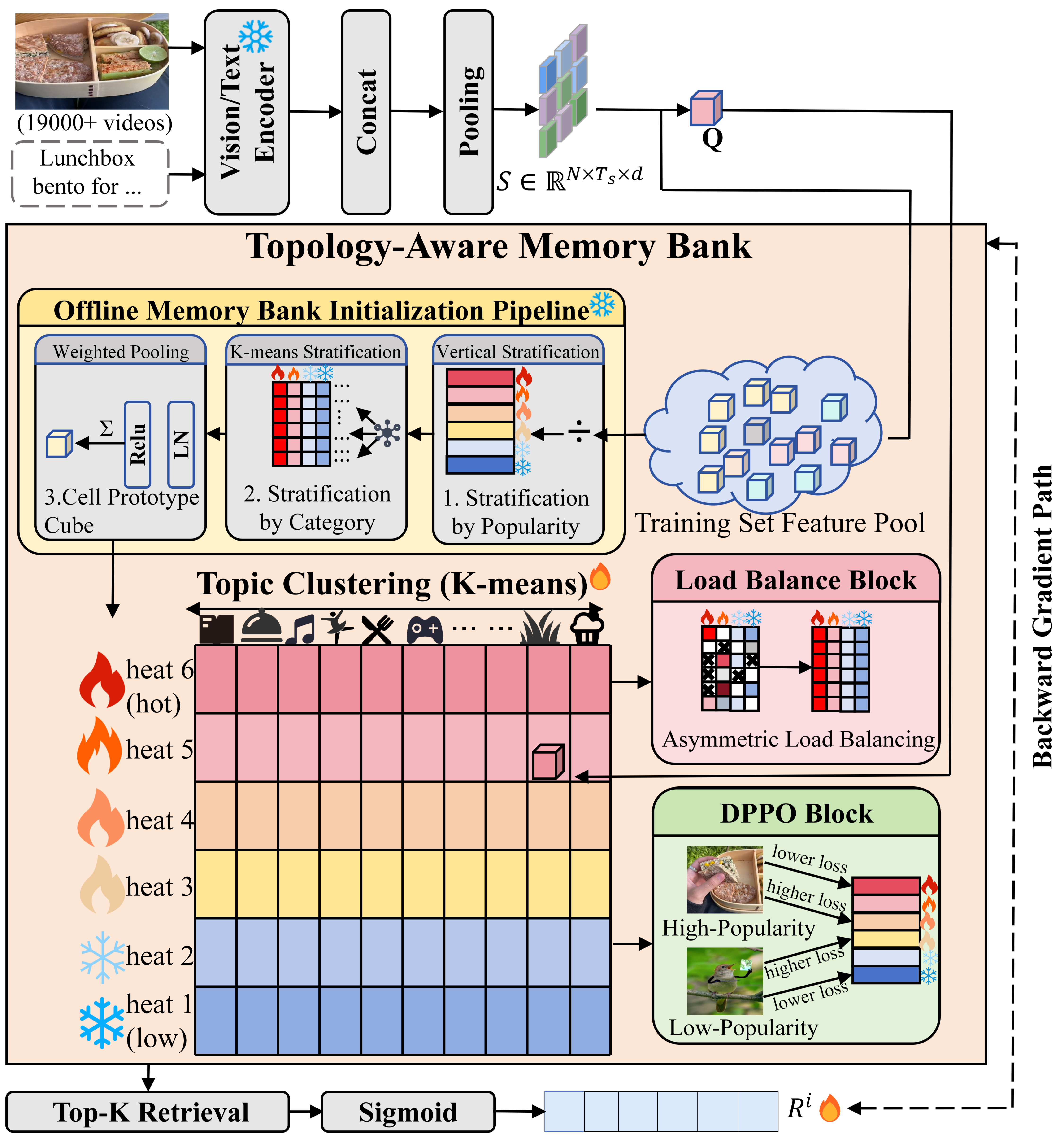}
  \caption{Spatial Enlargement workflow. The fused query enters a Topology-Aware Memory Bank initialized from training features, Prototype-Grid top-$K$ routing retrieves candidate memory slots, and Load Balance plus DPPO refinement stabilizes routing quality to produce retrieval feature $R_i$.}
  \Description{Spatial workflow from memory initialization and top-K routing to retrieval stabilization.}
  \label{fig:spatial_pipeline}
\end{figure}

\noindent\textbf{Offline Memory Bank Initialization Pipeline.}
We define memory as $\mathcal{M} \in \R^{P\times C\times d_m}$, where $P$ is the number of popularity partitions and $C$ is the number of topic clusters per partition. Each slot $\mathcal{M}_{p,c} \in \R^{d_m}$ is trainable. In our implementation, we set $P=6$ and $C=32$. For each training sample, we first construct the multimodal embedding $e_i=[V_i;T_i;U_i]$, then divide all training samples into $P$ popularity partitions by quantiles of ground-truth popularity $Y_i$. Within each partition $p$, we run K-Means with $C$ clusters on the corresponding embeddings and use cluster centers to initialize memory slots $\mathcal{M}_{p,c}$. The offline cluster assignment is only used for initialization; during training, sample-to-slot assignments (i.e., topology relations) are dynamically updated each iteration by current routing scores and top-$K$ selection, rather than frozen after initialization.

\noindent\textbf{Prototype-Grid Routing and Top-$K$ Retrieval.}
Given fused query $q_i$, we compute
\begin{equation}
  q_i'=\tanh(\mathrm{LN}(W_q q_i)),\qquad
  S_{i,p,c}=\frac{{q_i'}^\top\mathcal{M}_{p,c}}{\sqrt{d_m}}.
\end{equation}
With temperature $\tau$, sparse routing first computes the top-$K$ gate
weights as $\pi_i=\mathrm{TopK}(\mathrm{softmax}(S_i/\tau),K)$,
and retrieval augmentation is
\begin{equation}
  z_i^{\text{aug}} = \sum_{(p,c) \in \Omega_K}\pi_{i,p,c}\mathcal{M}_{p,c}.
\end{equation}
We set $K=5$ in all experiments. We also aggregate popularity context $c_i^{\text{pop}}=\sum_{(p,c) \in \Omega_K}\pi_{i,p,c}\mu_{p,c}$, where $\mu_{p,c}$ is the popularity centroid of slot $(p,c)$. Instead of fixing $\tau$, we treat routing temperature as a learnable parameter (with positivity constraint) and update it by backpropagation from prediction error in $\mathcal{L}_{\text{total}}$. For hard top-$K$ selection, gradients are propagated with a straight-through estimator. This routing stage corresponds to the Prototype-Grid and top-$K$ retrieval blocks in Figure~\ref{fig:spatial_pipeline}.

\noindent\textbf{Load Balance Block.}
To mitigate memory-slot collapse, we impose asymmetric priors: a power-law prior along popularity partitions and a uniform prior along topic clusters. The balancing loss is
\begin{equation}
  \mathcal{L}_{\text{balance}}=\alpha\,\mathrm{KL}(p^{(h)}\Vert p^{(h)}_{\text{prior}})+\beta\,\mathrm{KL}(p^{(t)}\Vert p^{(t)}_{\text{prior}}).
\end{equation}

\noindent\textbf{DPPO Block.}
Spearman's rank correlation coefficient (SRC), a core popularity prediction metric, emphasizes ranking consistency rather than absolute regression error. Regression-only supervision cannot sufficiently model pairwise ordering relations, so we introduce Dynamic Pairwise Preference Optimization (DPPO) to align Spatial Enlargement routing with popularity-aware ranking supervision, inspired by pairwise ranking and preference optimization principles~\cite{rendle2009bpr,rafailov2023dpo}.

For pair samples $(\mathcal{S}^+,\mathcal{S}^-)$ satisfying $Y^+-Y^->m$, we encourage higher routing confidence for the more popular sample:
\begin{equation}
  \mathcal{L}_{\text{pref}}=-\log\sigma\!\left(\gamma\sum_{(p,c)}\left[\log\pi^+_{p,c}-\log\pi^-_{p,c}\right]\right).
\end{equation}
This ranking-aware supervision aligns retrieval behavior with target ranking metrics.

\noindent\textbf{Spatial Optimization.}
Memory slots $\mathcal{M}_{p,c}$ are treated as trainable parameters during end-to-end learning. Gradients from regression, preference, and balancing losses are backpropagated through retrieval aggregation to jointly update query projection, routing temperature, and memory slots, enabling continuous refinement of stored semantic and popularity knowledge.
\subsection{Popularity Prediction Procedure}
\method{} follows a three-stage design aligned with Figure~\ref{fig:overview}: (i) Temporal Enlargement (Sec.~3.2) for long-context visual modeling, (ii) Spatial Enlargement (Sec.~3.3) for topology-aware retrieval augmentation, and (iii) Popularity Prediction for final inference. The first two stages are presented in Sec.~3.2 and Sec.~3.3, respectively. We next formalize the third stage and then analyze the overall complexity of the complete framework.

\noindent\textbf{Bi-directional Cross-Attention and Predictor.}
To model visual--textual interaction under noisy multimodal inputs, we employ bi-directional cross-attention following CBAN-style interaction~\cite{cheung2022crossmodal}:
\begin{equation}
  T_i^{\star}=\mathrm{Att}_{t}(V_i,T_i,T_i),\qquad
  V_i^{\star}=\mathrm{Att}_{v}(T_i,V_i,V_i).
\end{equation}
We then fuse retrieval features through $\mathbf{R}_i=\mathrm{ReLU}\!\left(W_R[z_i^{\text{aug}};c_i^{\text{pop}}]\right)$.

\noindent\textbf{Prediction Objective.}
The final predictor input is written as $H_i=[V_i^{\star};T_i^{\star};U_i;\mathbf{R}_i;V_i^{\star}\odot \mathbf{R}_i;T_i^{\star}\odot \mathbf{R}_i]$.
Popularity is regressed by a lightweight MLP head. The total objective is
\begin{equation}
  \mathcal{L}_{\text{total}}=\mathcal{L}_{\text{reg}}+\lambda_{\text{pref}}\mathcal{L}_{\text{pref}}+\lambda_{\text{bal}}\mathcal{L}_{\text{balance}},
\end{equation}
where $\mathcal{L}_{\text{reg}}$ is Huber loss, and $\lambda_{\text{pref}},\lambda_{\text{bal}}$ control ranking and balancing terms.

\noindent\textbf{Complexity Analysis.}
\begin{table}[H]
  \centering
  \small
  \caption{Simplified per-sample complexity comparison (assuming $PC\ll N$ and $\bar{w}\ll T_w$).}
  \label{tab:complexity}
  \setlength{\tabcolsep}{2pt}
  \begin{tabularx}{\columnwidth}{>{\raggedright\arraybackslash}p{0.27\columnwidth}>{\centering\arraybackslash}X>{\centering\arraybackslash}X}
    \toprule
    Method & Time Complexity (lower is better) & Space Complexity \\
    \midrule
    Traditional methods & $\mathcal{O}(TT_w + N)$ & $\mathcal{O}(N)$ \\
    \method{} (ours) & \textbf{$\mathcal{O}(T\bar{w} + PC)$} & \textbf{$\mathcal{O}(PC)$} \\
    \bottomrule
  \end{tabularx}
\end{table}

For Temporal Enlargement, traditional popularity prediction temporal modeling often relies on short-window dense attention, with sequence-level time complexity $\mathcal{O}(TT_wd_a)$ ($T_w\ll T$), which limits long-range context coverage. Our temporal stage combines the SSM Block $\mathcal{O}(Td_h)$ and the Sparse Attention Block $\mathcal{O}(T\bar{w}d_a)$ under the Frame Scoring Module, yielding total complexity $\mathcal{O}(Td_h+T\bar{w}d_a)$.

For Spatial Enlargement, traditional retrieval pipelines search top-$K$ neighbors over all $N$ historical videos, leading to $\mathcal{O}(Nd_m+Kd_m)$ retrieval time and $\mathcal{O}(Nd_m)$ memory storage. Our Topology-Aware Memory Bank with Prototype-Grid routing computes scores on only $P\times C$ slots and then performs sparse top-$K$ aggregation, requiring $\mathcal{O}(PCd_m+Kd_m)$ retrieval time and $\mathcal{O}(PCd_m)$ storage. This yields substantially better scalability and memory efficiency in large-scale deployment when $PC\ll N$.

In practical settings where $PC\ll N$ and $\bar{w}\ll T_w$, \method{} reduces both dominant computational complexity and memory overhead relative to traditional pipelines.
\section{Experiments}
We now present experimental results to validate \method{} and answer the following research questions:
\begin{itemize}
  \item \textbf{RQ1:} How does \method{} compare with strong baselines on three datasets?
  \item \textbf{RQ2:} How much does each key module contribute to overall performance?
  \item \textbf{RQ3:} How sensitive is \method{} to key hyperparameters?
  \item \textbf{RQ4:} How do Temporal Enlargement (RQ4-T) and topology-aware memory (RQ4-S) improve temporal robustness and spatial cold-start generalization?
\end{itemize}
\subsection{Experimental Settings}
\noindent\textbf{Dataset.}
To analyze the effectiveness of \method{}, we select three real-world micro-video datasets from various online video platforms: MicroLens~\cite{ni2023microlens}, SMPD-video~\cite{wu2025smpv}, and Informs (\url{https://connect.informs.org/}). The three datasets contain 19,307, 4,000, and 1,846 videos, respectively. In our experiments, MicroLens is split with a ratio of $7{:}1{:}2$, SMPD-video is split with a ratio of $8{:}1{:}1$, and Informs follows its official split. For visual processing, we extract 30 frames per video on MicroLens, and 100 frames per video on SMPD-video and Informs. Here, $d_v$ and $d_t$ denote the dimensions of visual and textual features, respectively.

\noindent\textbf{Baselines.}
To evaluate the superiority of \method{}, we compare against representative baselines implemented in our project from two groups:
\begin{itemize}
  \item Feature-engineering methods: SVR~\cite{khosla2014makes}, HyFea~\cite{lai2020hyfea}, and MFTM~\cite{hsu2023mftm}.
  \item Multimodal fusion methods: Contextual-LSTM (CLSTM)~\cite{ghosh2016clstm}, TMALL~\cite{chen2016micro}, MASSL~\cite{zhang2022massl}, CBAN~\cite{cheung2022crossmodal}, HMMVED~\cite{xie2023hmmved}, MQMC~\cite{du2023mqmc}, MMRA~\cite{zhong2024mmra}, and ICPF~\cite{cheng2024icpf}.
\end{itemize}

\noindent\textbf{Metrics.}
Following prior works~\cite{chen2016micro,zhong2024mmra}, we use three widely adopted metrics: normalized mean squared error (nMSE), mean absolute error (MAE), and Spearman's rank correlation coefficient (SRC).

\noindent\textbf{Model implementation.}
During retrieval, we use CLIP ViT-L/14~\cite{radford2021clip} as the image encoder and UAE-Large-V1 (WhereIsAI) as the text encoder. The textual feature dimension is set to 1024. We optimize the model using AdamW~\cite{loshchilov2017adamw} with a learning rate of $2\times 10^{-4}$, while the encoder learning rate is set to $8\times 10^{-6}$. The model is trained for 50 epochs with a batch size of 32.
\subsection{Main Results (RQ1)}
\begin{table*}[!t]
  \centering
  \footnotesize
  \setlength{\tabcolsep}{2pt}
  \caption{Performance comparison on three real-world datasets. The best results are in bold font and the second underlined. Lower values of nMSE and MAE, and higher values of SRC, indicate better performance.}
  \label{tab:main}
  \begin{tabularx}{\textwidth}{l|l|>{\centering\arraybackslash}X>{\centering\arraybackslash}X>{\centering\arraybackslash}X|>{\centering\arraybackslash}X>{\centering\arraybackslash}X>{\centering\arraybackslash}X>{\centering\arraybackslash}X>{\centering\arraybackslash}X>{\centering\arraybackslash}X>{\centering\arraybackslash}X>{\centering\arraybackslash}X|>{\columncolor{gray!15}\centering\arraybackslash}X|>{\centering\arraybackslash}X|>{\centering\arraybackslash}X}
    \hline
    Dataset & Metric & SVR & HyFea & MFTM & CLSTM & TMALL & MASSL & CBAN & HMMVED & MQMC & MMRA & ICPF & \textbf{\method} & Improv. & p-val. \\
    \hline
    \multirow{3}{*}{MicroLens} & nMSE & 0.7829 & 0.7481 & 0.7826 & 1.0832 & 0.7567 & 1.1090 & 0.9906 & 1.1912 & 2.4157 & 0.7512 & \underline{0.7305} & \textbf{0.7148} & 2.15\%$\downarrow$ & \mbox{3.79e-2} \\
    & MAE & 1.0899 & 1.0846 & 1.1111 & 1.1830 & \underline{1.0830} & 1.3206 & 1.2542 & 1.1326 & 1.2369 & 1.1887 & 1.1088 & \textbf{1.0482} & 3.21\%$\downarrow$ & \mbox{8.52e-3} \\
    & SRC & 0.4796 & 0.4974 & 0.4662 & 0.3332 & 0.4871 & 0.0120 & 0.1797 & 0.4311 & 0.4002 & 0.4951 & \underline{0.5102} & \textbf{0.5346} & 4.78\%$\uparrow$ & \mbox{3.22e-4} \\
    \hline
    \multirow{3}{*}{SMPD-video} & nMSE & 0.8102 & 0.8932 & 0.8822 & 0.8477 & 1.2025 & 1.1334 & 0.8777 & 0.9254 & 6.1899 & 0.9144 & \underline{0.7611} & \textbf{0.4887} & 35.79\%$\downarrow$ & \mbox{9.90e-7} \\
    & MAE & 1.9559 & 2.0711 & 2.0182 & 1.9628 & 2.3752 & 2.3037 & 2.0011 & 2.0455 & 1.9343 & 2.0345 & \underline{1.8358} & \textbf{1.4349} & 21.84\%$\downarrow$ & \mbox{4.20e-7} \\
    & SRC & 0.4455 & 0.3416 & 0.3734 & 0.4187 & 0.2137 & 0.3788 & 0.3951 & 0.4331 & 0.4274 & 0.3849 & \underline{0.4980} & \textbf{0.6794} & 36.42\%$\uparrow$ & \mbox{3.08e-6} \\
    \hline
    \multirow{3}{*}{Informs} & nMSE & 0.7243 & 0.8185 & \underline{0.6803} & 0.8610 & 6.9960 & 1.2386 & 0.8482 & 1.0762 & 0.8524 & 0.8391 & 0.9551 & \textbf{0.5806} & 14.66\%$\downarrow$ & \mbox{8.00e-6} \\
    & MAE & 0.0997 & 0.1042 & \underline{0.0966} & 0.1098 & 0.3115 & 0.1343 & 0.1093 & 0.1260 & 0.1116 & 0.1084 & 0.1153 & \textbf{0.0876} & 9.32\%$\downarrow$ & \mbox{3.00e-5} \\
    & SRC & 0.5720 & 0.4823 & \underline{0.6357} & 0.4987 & 0.0674 & 0.0596 & 0.5064 & 0.0564 & 0.3941 & 0.5052 & 0.4985 & \textbf{0.6460} & 1.62\%$\uparrow$ & \mbox{5.06e-3} \\
    \hline
  \end{tabularx}
\end{table*}

We compare \method{} with 11 strong baselines on three datasets. Quantitative results are reported in Table~\ref{tab:main}.

Table~\ref{tab:main} supports two key observations:
\begin{itemize}
  \item \textbf{(O1)} \method{} consistently outperforms all competitive baselines across all datasets and metrics. In particular, on SMPD-video, \method{} improves nMSE, MAE, and SRC by 35.79\%$\downarrow$, 21.84\%$\downarrow$, and 36.42\%$\uparrow$, respectively, compared with the strongest baseline. Similar gains are also observed on MicroLens and Informs. Additionally, we retrain both \method{} and the best-performing baselines five times to assess the p-value. These results validate that jointly optimized Temporal Enlargement and Spatial Enlargement can provide more accurate popularity prediction and stronger ranking consistency.
  \item \textbf{(O2)} \method{} significantly outperforms both feature-engineering methods and deep multimodal methods (including retrieval-augmented MMRA and ICPF). We attribute this advantage to our joint spatio-temporal enlargement design: Temporal Enlargement captures sparse long-range highlight cues through the Frame Scoring Module, SSM Block, and Sparse Attention Block, while Spatial Enlargement provides denoised and popularity-aware retrieval context via the Topology-Aware Memory Bank, Load Balance Block, and DPPO Block. In contrast, methods relying on short temporal windows, flat memory retrieval, or static prompting strategies are less effective at modeling cross-video contextual correlations for popularity prediction.
\end{itemize}
These cross-group gains indicate that improvements from Temporal Enlargement and Spatial Enlargement are interaction-driven rather than a simple linear combination of two independent modules.
\subsection{Ablation Study (RQ2)}
We conduct extensive ablation experiments to evaluate the impact of each critical component in \method{}. Following our architecture design, we analyze both Temporal Enlargement and Spatial Enlargement on MicroLens and SMPD-video. The detailed results are reported in Table~\ref{tab:ablation}.

\begin{table}[!t]
  \centering
  \footnotesize
  \caption{Ablation study of \method{} on key components. Lower values of nMSE and MAE, and higher values of SRC, indicate better performance.}
  \label{tab:ablation}
  \setlength{\tabcolsep}{2pt}
  \arrayrulecolor{black}
  \begin{tabularx}{\columnwidth}{l|>{\raggedright\arraybackslash}Xccc|ccc}
    \hline
    \multirow{2}{*}{Module} & \multirow{2}{*}{Variant} & \multicolumn{3}{c|}{MicroLens} & \multicolumn{3}{c}{SMPD-video} \\
    \cline{3-8}
    &  & nMSE & MAE & SRC & nMSE & MAE & SRC \\
    \hline
    \textbf{\method{}} & \textbf{All} & \textbf{0.7114} & \textbf{1.0463} & \textbf{0.5346} & \textbf{0.4887} & \textbf{1.4349} & \textbf{0.6794} \\
    \hline
    \multirow{4}{*}{Temporal} & w/o Frame Scoring & 0.7420 & 1.0616 & 0.5227 & 0.5000 & 1.4715 & 0.6554 \\
    & w/o SSM Block & 0.7157 & 1.0490 & 0.5288 & 0.4941 & 1.4686 & 0.6571 \\
    & w/o Sparse Attention & 0.7686 & 1.0767 & 0.5015 & 0.4936 & 1.4582 & 0.6682 \\
    & w/o visual temporal & 1.0022 & 1.2563 & 0.0922 & 0.5298 & 1.5186 & 0.6370 \\
    \hline
    \multirow{4}{*}{Spatial} & Top-1 memory & 0.7352 & 1.0526 & 0.5239 & 0.5206 & 1.4824 & 0.6512 \\
    & w/o DPPO & 0.7161 & 1.0502 & 0.5282 & 0.4839 & 1.4464 & 0.6666 \\
    & w/o Load Balance & 0.7162 & 1.0489 & 0.5295 & 0.5507 & 1.5392 & 0.6428 \\
    & w/o PMB & 0.7331 & 1.0744 & 0.5156 & 0.5967 & 1.5730 & 0.6302 \\
    \hline
  \end{tabularx}
\end{table}

\begin{figure}[!b]
  \centering
  \includegraphics[width=0.98\linewidth]{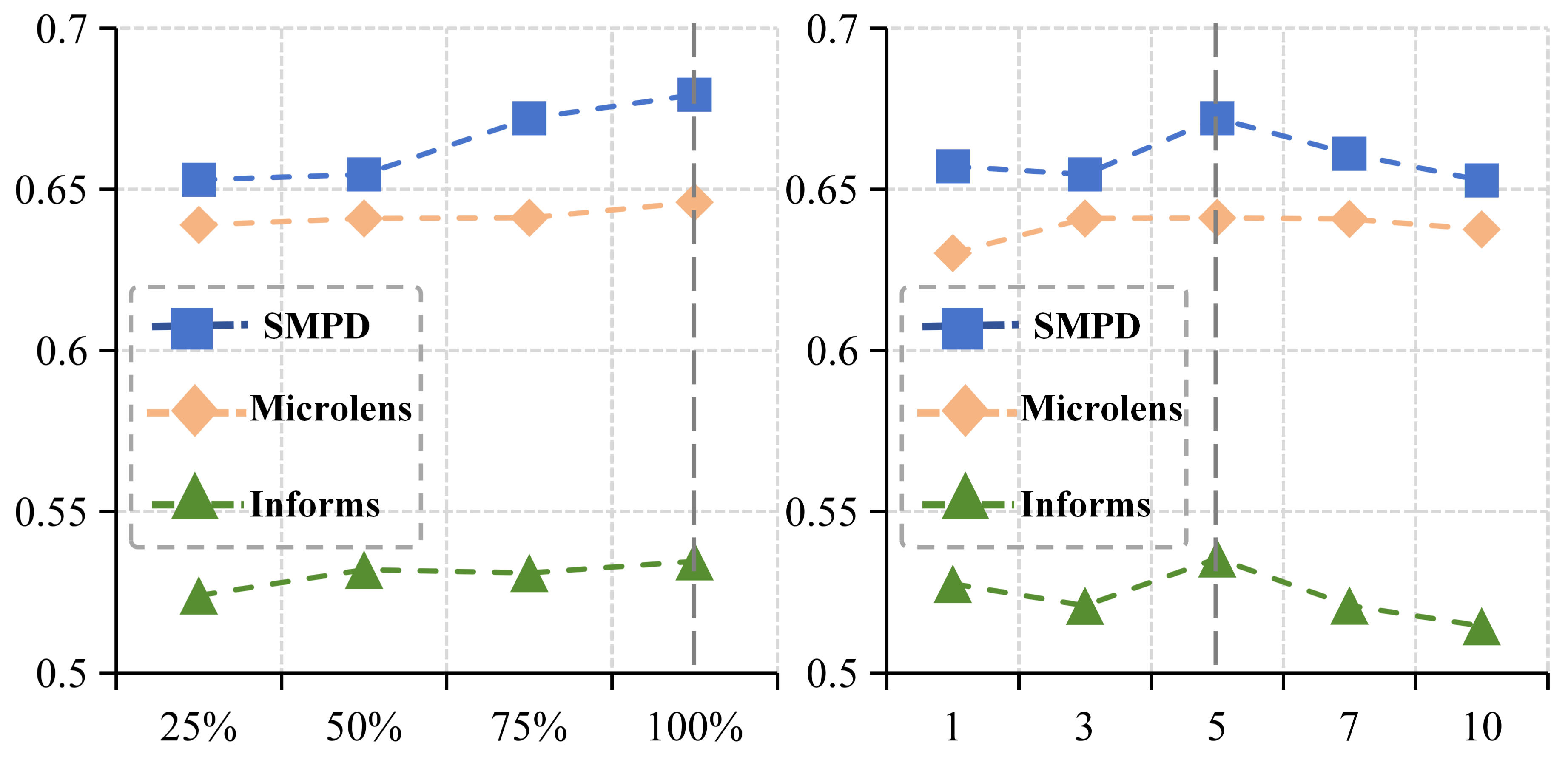}
  \caption{Sensitivity analysis of \method{} on key hyperparameters. Left: base window size in the adaptive highlight window of the Sparse Attention Block. Right: top-$K$ retrieval selection in the Topology-Aware Memory Bank.}
  \Description{Two-panel sensitivity figure on SMPD-video, MicroLens, and Informs. The left panel varies base window size from 25 percent to 100 percent for the adaptive highlight window. The right panel varies top-$K$ memory retrieval from 1 to 10 for the Topology-Aware Memory Bank.}
  \label{fig:sensitivity}
\end{figure}

Table~\ref{tab:ablation} supports two observations:
\begin{itemize}
  \item \textbf{Effect of Temporal Enlargement.} To assess the temporal stage, we design four variants: (1) w/o Frame Scoring, removing the Frame Scoring Module; (2) w/o SSM Block, removing the SSM Block; (3) w/o Sparse Attention, removing the Sparse Attention Block; and (4) w/o visual temporal, removing the temporal visual branch. In Table~\ref{tab:ablation}, all four variants show clear performance drops on both datasets, and removing the temporal visual branch causes the largest degradation. These results verify that the Frame Scoring, SSM, and Sparse Attention blocks are all necessary for effective long-context temporal modeling.
  \item \textbf{Effect of Spatial Enlargement.} To assess the spatial stage, we design four variants: (1) Top-1 memory, replacing Prototype-Grid top-$K$ routing with single-slot retrieval; (2) w/o DPPO, removing the DPPO Block; (3) w/o Load Balance, removing the Load Balance Block; and (4) w/o PMB, reducing memory-bank capacity. As shown in Table~\ref{tab:ablation}, all spatial variants underperform the full model. Specifically, Top-1 memory weakens retrieval coverage, w/o DPPO reduces ranking consistency, w/o Load Balance harms routing stability, and reduced memory-bank capacity causes clear overall decline. These findings validate the effectiveness of topology-aware structured memory routing and DPPO-aware optimization.
\end{itemize}
Notably, both temporal-side and spatial-side ablations lead to synchronized degradation across all three metrics on both datasets, indicating that each branch depends on the other and that the full-model gain is not a simple additive effect.
\subsection{Sensitivity (RQ3)}
Figure~\ref{fig:sensitivity} presents the sensitivity analysis of two key hyperparameters in \method{}.

\noindent\textbf{Base window size for the adaptive highlight window.}
The left panel of Figure~\ref{fig:sensitivity} shows the effect of base window size on three datasets.
Performance on SMPD-video and Informs generally improves as the base window size increases, whereas MicroLens remains relatively stable across settings.
The 100\% setting achieves the best or near-best results across all datasets.
This pattern suggests that a larger base window preserves broader temporal context, while the dynamic mechanism can still reduce the effective window for less informative frames.
Accordingly, we set the base window size to 100\%.

\noindent\textbf{Top-$K$ memory selection in the Topology-Aware Memory Bank.}
The right panel of Figure~\ref{fig:sensitivity} reports the impact of top-$K$ memory retrieval.
When $K=1$, the model relies on only one memory slot and the retrieved information is often insufficient.
As $K$ increases, performance improves and reaches the best balance at $K=5$.
When $K$ becomes larger (e.g., $K=10$), performance drops on multiple datasets, likely due to introducing irrelevant memory noise.
Therefore, we adopt top-$K=5$ as the default configuration.
\subsection{Innovation-Oriented Analysis (RQ4)}
We analyze RQ4 directly from our two core innovations: temporal-side effectiveness of Temporal Enlargement (RQ4-T) and spatial-side effectiveness of the Topology-Aware Memory Bank (RQ4-S).

\begin{figure}[t]
  \centering
  \includegraphics[width=0.98\linewidth]{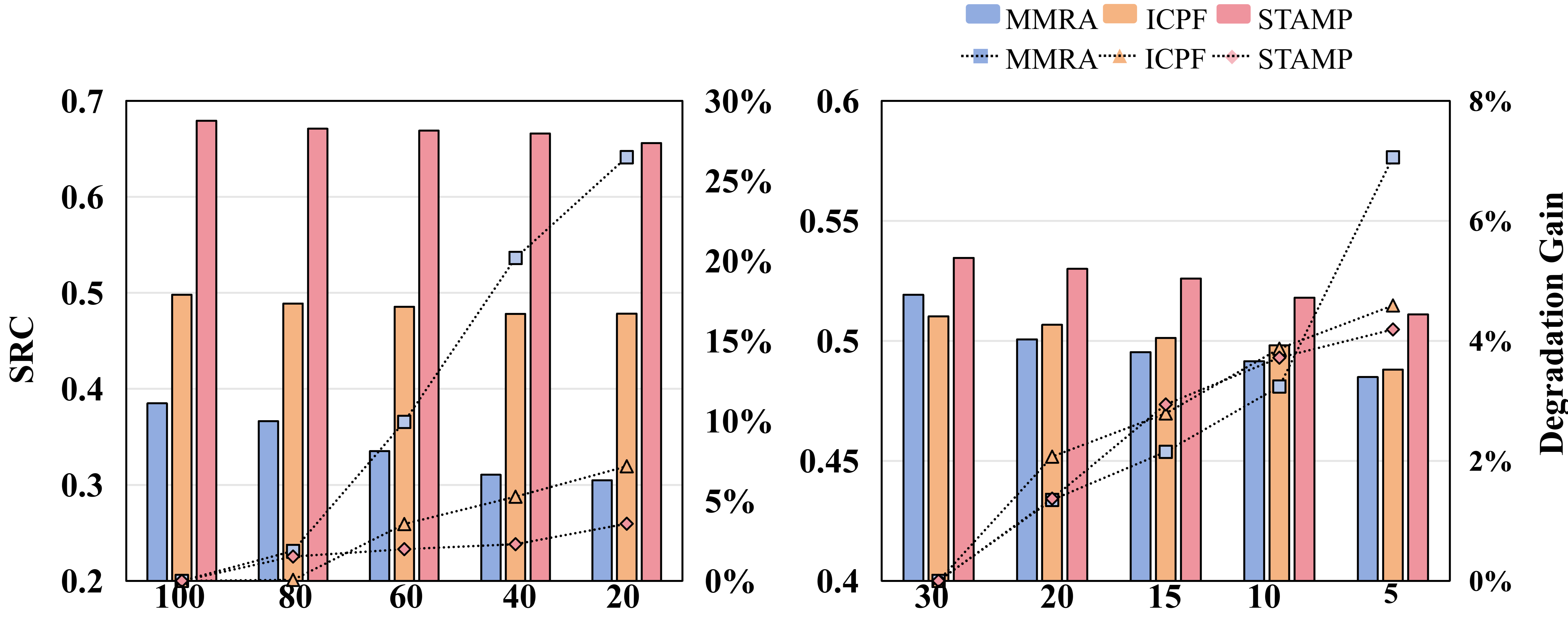}
  \makebox[\linewidth][c]{(a) SMPD-video\hspace{0.20\linewidth}(b) MicroLens}
  \caption{Temporal-side robustness under different training-set proportions. The bars represent SRC values and the polylines denote the performance degradation gain.}
  \Description{Two-panel temporal-robustness figure. Panel (a) shows SMPD-video and panel (b) shows MicroLens. Bars present SRC values for MMRA, ICPF, and STAP across training-set proportions, while polylines show degradation gain trends.}
  \label{fig:robustness}
\end{figure}

\noindent\textbf{RQ4-T: Temporal Enlargement.}
To assess temporal robustness under limited supervision, we compare \method{} with competitive retrieval-augmented baselines across varying training-set proportions on two benchmarks widely used in recent popularity prediction studies~\cite{wu2025smpv,ye2025mvp}. As shown in Figure~\ref{fig:robustness}, \method{} generally achieves higher SRC values than MMRA and ICPF in the evaluated settings~\cite{zhong2024mmra,cheng2024icpf}. In addition, \method{} exhibits consistently smaller performance degradation as training data decreases, indicating stronger temporal robustness under data scarcity. \textbf{These temporal-side results are consistent with the objective of Temporal Enlargement and indicate more stable generalization under limited supervision, attributable to the coordinated Frame Scoring, SSM, and Sparse Attention blocks that preserve sparse long-range cues and mitigate sensitivity to supervision sparsity.}

\begin{table}[!t]
  \centering
  \footnotesize
  \caption{Macro-level quantitative comparison on retrieved cold-start subsets for RQ4-S. Lower MAE and higher SRC indicate better performance.}
  \label{tab:rq4s_coldstart}
  \setlength{\tabcolsep}{2.5pt}
  \begin{tabularx}{\columnwidth}{>{\raggedright\arraybackslash}X>{\raggedright\arraybackslash}X>{\centering\arraybackslash}p{0.16\columnwidth}>{\centering\arraybackslash}p{0.16\columnwidth}}
    \toprule
    Dataset & Method & MAE$\downarrow$ & SRC$\uparrow$ \\
    \midrule
    \multirow{3}{*}{SMPD-video} & \method{} (Top-5) & \textbf{1.8953} & \textbf{0.5321} \\
    & ICPF (Top-10) & 2.0201 & 0.3240 \\
    & MMRA (Top-10) & 1.9323 & 0.4104 \\
    \midrule
    \multirow{3}{*}{Informs} & \method{} (Top-5) & \textbf{0.1272} & \textbf{0.2813} \\
    & MMRA (Top-10) & 0.1324 & 0.1885 \\
    & ICPF (Top-10) & 0.1501 & 0.2454 \\
    \bottomrule
  \end{tabularx}
\end{table}

\begin{figure}[!t]
  \centering
  \vspace{-2pt}
  \includegraphics[width=0.72\linewidth]{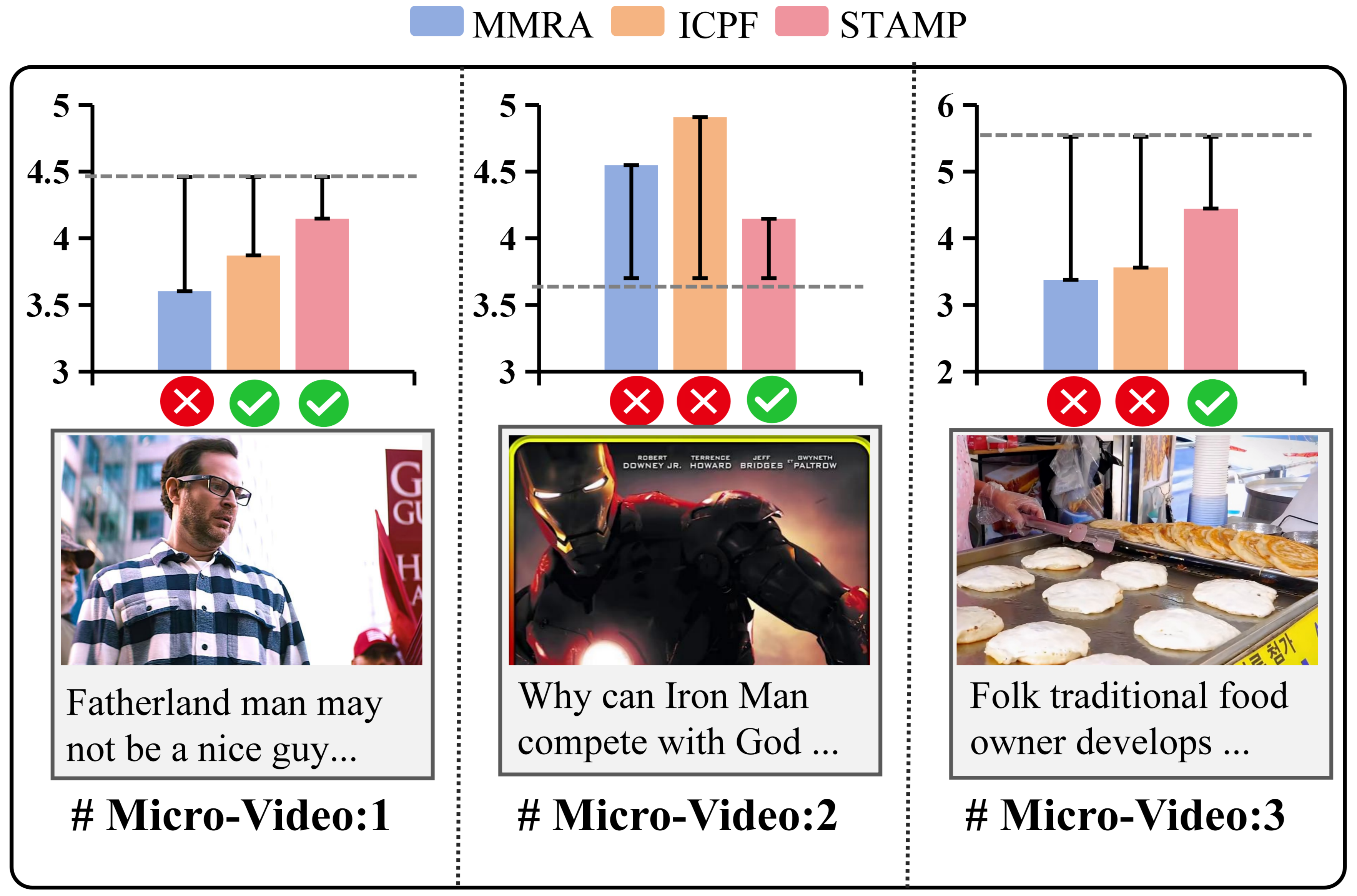}
  \vspace{-4pt}
  \caption{Spatial-memory retrieval quality on representative cold-start cases (users with $\leq 3$ published videos). Top: predictions of MMRA, ICPF, and \method{} against ground truth (gray dashed line). Bottom: corresponding video and text snippets.}
  \Description{Three representative cold-start cases comparing predictions from MMRA, ICPF, and STAP with ground truth. Red-cross and green-check markers denote errors above or below 20 percent, and the bottom row shows corresponding video-text snippets.}
  \label{fig:case}
\end{figure}

\noindent\textbf{RQ4-S: Spatial Memory Bank.}
To assess spatial robustness under cold-start settings, we conduct a macro-level quantitative evaluation on retrieved cold-start subsets from SMPD-video and Informs, using MAE and SRC as metrics. As shown in Table~\ref{tab:rq4s_coldstart}, \method{} (Top-5) achieves the best MAE and SRC on both datasets, indicating that topology-aware memory routing provides more reliable support under cold-start conditions than flat retrieval baselines~\cite{zhong2024mmra,cheng2024icpf}. Figure~\ref{fig:case} presents representative cold-start cases consistent with this macro-level trend. \textbf{These spatial-side results are consistent with the objective of Spatial Enlargement and indicate stronger cold-start generalization, attributable to the Topology-Aware Memory Bank with Prototype-Grid routing, Load Balance Block, and DPPO Block, which improves retrieval relevance and suppresses memory noise.}

\vspace{-6pt}

\section{Conclusion}
\vspace{-2pt}
In this work, we propose \method{}, a unified spatio-temporal enlargement framework for micro-video popularity prediction. \method{} integrates a Temporal Enlargement stage (Frame Scoring Module, SSM Block, Sparse Attention Block, and Gated Fusion Mechanism), a Spatial Enlargement stage (Topology-Aware Memory Bank with Prototype-Grid routing, Load Balance Block, and DPPO Block), and a popularity prediction stage with bi-directional cross-attention and a predictor. Extensive experiments on three real-world datasets validate the effectiveness of \method{}. In future work, we plan to extend \method{} to broader scenarios, including multimodal recommendation, multimodal graph learning, and other multimodal forecasting tasks.

\bibliographystyle{ACM-Reference-Format}
\bibliography{main}

\clearpage
\input{main_append_body}

\end{document}

%% file: main_append_body.tex
\appendix
\setcounter{table}{0}
\setcounter{figure}{0}
\setcounter{algorithm}{0}
\renewcommand{\thetable}{S\arabic{table}}
\renewcommand{\thefigure}{S\arabic{figure}}
\renewcommand{\thealgorithm}{S\arabic{algorithm}}
\providecommand*{\theHtable}{\thetable}
\providecommand*{\theHfigure}{\thefigure}
\providecommand*{\theHalgorithm}{\thealgorithm}
\renewcommand*{\theHtable}{appendix.table.\arabic{table}}
\renewcommand*{\theHfigure}{appendix.figure.\arabic{figure}}
\renewcommand*{\theHalgorithm}{appendix.algorithm.\arabic{algorithm}}
\raggedbottom
\setlength{\textfloatsep}{3pt plus 1pt minus 1pt}
\setlength{\dbltextfloatsep}{3pt plus 1pt minus 1pt}
\setlength{\floatsep}{3pt plus 1pt minus 1pt}
\setlength{\dblfloatsep}{3pt plus 1pt minus 1pt}
\setlength{\intextsep}{3pt plus 1pt minus 1pt}
\setlength{\abovecaptionskip}{2pt}
\setlength{\belowcaptionskip}{0pt}
\setlength{\abovedisplayskip}{4pt plus 1pt minus 1pt}
\setlength{\belowdisplayskip}{4pt plus 1pt minus 1pt}
\setlength{\abovedisplayshortskip}{2pt plus 1pt minus 1pt}
\setlength{\belowdisplayshortskip}{2pt plus 1pt minus 1pt}
\setcounter{topnumber}{4}
\setcounter{bottomnumber}{4}
\setcounter{dbltopnumber}{4}
\setcounter{totalnumber}{8}
\renewcommand{\topfraction}{0.95}
\renewcommand{\dbltopfraction}{0.95}
\renewcommand{\bottomfraction}{0.95}
\renewcommand{\textfraction}{0.05}
\renewcommand{\floatpagefraction}{0.85}
\renewcommand{\dblfloatpagefraction}{0.85}
\providecommand{\promptbox}[1]{%
  \noindent\fcolorbox{black!12}{gray!6}{%
    \parbox{\dimexpr\linewidth-2\fboxsep-2\fboxrule\relax}{#1}}%
}

\begin{center}
  {\LARGE\bfseries APPENDIX}
\end{center}
\smallskip

\noindent The appendix is organized as follows. \S A presents complete implementation details and reproducibility settings. \S B reports empirical efficiency and system overhead. \S C analyzes spatial memory behavior and bank-scale selection. \S D gives temporal frame-level scoring visualization and interpretation.

\section{Implementation Details}

\subsection{Feature Extraction}
We provide the exact preprocessing pipeline used for visual, textual, and optional audio modalities. The full pipeline is deterministic except for data-loader shuffling.

\noindent\textbf{Visual branch: CLIP ViT-L/14 input.}
\begin{itemize}[topsep=2pt,itemsep=1pt,parsep=0pt,partopsep=0pt]
  \item \textbf{Frame sampling:} we use deterministic uniform sampling with target length $T$ ($T{=}31$ for MicroLens and $T{=}100$ for SMPD-video/Informs); if a video is shorter than $T$, the last valid frame is repeated.
  \item \textbf{Frame encoding:} sampled frames are normalized with CLIP preprocessing and encoded by CLIP ViT-L/14; temporal order is preserved for the downstream temporal module.
\end{itemize}

\noindent\textbf{Text branch: dataset-specific sources.}
\begin{itemize}[topsep=2pt,itemsep=1pt,parsep=0pt,partopsep=0pt]
  \item \textbf{SMPD-video:} text features are constructed from \texttt{post\_content} and \texttt{post\_suggested\_words}, which are converted into hashtag-style tokens.
  \item \textbf{Informs:} the original videos do not provide native text descriptions, so textual features are generated by LLaVA-NeXT-Video-7B-hf from a fixed prompt set and combined with regex-extracted hashtags and mentions. The fixed prompt set is:
  \par\smallskip
  \promptbox{%
    \footnotesize\raggedright\setlength{\parskip}{2.5pt}\itshape
    ``Can you analyze the video in terms of its content interestingness, including emotional appeal, engagement strategies, pacing, and uniqueness?''\par
    ``How do the scene transitions affect pacing, engagement, and narrative flow?''\par
    ``How do camera movements like panning, zooming, and tracking enhance storytelling, emotional tone, and immersion?''\par
    ``What is the central plot conflict, and how does it affect character development and viewer engagement?''\par
    ``How would you evaluate the quality of presentation, including visuals, audio, and production techniques?''\par
    ``What makes the video entertaining, and how do engaging moments and visual storytelling contribute to its popularity?''\par
    ``How does the genre influence its initial and long-term popularity?''\par
    ``Finally, how does the emotional tone conveyed through visuals, music, and narration impact viewer connection?''%
  }
  \item \textbf{MicroLens:} text features concatenate title, hashtags, and category before encoding.
  \item \textbf{Tokenization behavior:} the pipeline relies on the internal tokenizers of SentenceTransformer and AnglE; no explicit truncation length is specified in the preprocessing or model code.
\end{itemize}

\subsection{Algorithm Pseudocode}
Algorithm~\ref{alg:ssm_dynamic_scan} describes dynamic temporal scanning in the SSM pathway, while Algorithm~\ref{alg:routing_update} describes topology-aware routing and memory-slot updates.

\begin{algorithm}[t]
  \caption{Dynamic SSM scan in \method{}, PyTorch style}
  \label{alg:ssm_dynamic_scan}
  \begin{algorithmic}[1]
    \Require frame features $\mathbf{X}\in\R^{B\times T\times d_h}$, frame scores $\mathbf{s}\in\R^{B\times T}$
    \Require base step $\Delta_0$, control factor $\alpha$, state params $(\Lambda,\mathbf{B},\mathbf{C})$
    \State $\hat{\mathbf{s}}\gets\operatorname{sigmoid}(\mathbf{s})$ \Comment{score in $(0,1)$}
    \State $\mathbf{\Delta}\gets\Delta_0\cdot(1+\alpha\cdot(1-\hat{\mathbf{s}}))$
    \State $\mathbf{h}_0\gets\mathbf{0}$
    \For{$t=1$ to $T$}
      \State $\mathbf{A}_t\gets\operatorname{exp}(-\mathbf{\Delta}_{:,t}\odot\Lambda)$
      \State $\mathbf{h}_t\gets\mathbf{A}_t\odot\mathbf{h}_{t-1}+\mathbf{B}\mathbf{X}_{:,t,:}$
      \State $\mathbf{Y}^{ssm}_{:,t,:}\gets\mathbf{C}\mathbf{h}_t$
    \EndFor
    \State \Return $\mathbf{Y}^{ssm}$
  \end{algorithmic}
\end{algorithm}

\begin{algorithm}[t]
  \caption{Topology-aware routing and memory-slot update, PyTorch style}
  \label{alg:routing_update}
  \begin{algorithmic}[1]
    \Require query $\mathbf{q}\in\R^{B\times d_m}$, memory bank $\mathbf{M}\in\R^{P\times C\times d_m}$
    \Require popularity centroids $\boldsymbol{\mu}\in\R^{P\times C}$, routing temperature $\tau$, top-$K$
    \State $\mathbf{S}\gets\operatorname{einsum}(\text{bd,pcd->bpc},\mathbf{q},\mathbf{M})$
    \State $\boldsymbol{\pi}\gets\operatorname{softmax}(\mathbf{S}/\tau)$
    \State $(\boldsymbol{\pi}_K,\Omega_K)\gets\operatorname{TopK}(\boldsymbol{\pi},K)$
    \State $\mathbf{r}\gets\sum\limits_{(p,c)\in\Omega_K}\boldsymbol{\pi}_{K,p,c}\cdot\mathbf{M}_{p,c}$
    \State $\mathbf{c}^{pop}\gets\sum\limits_{(p,c)\in\Omega_K}\boldsymbol{\pi}_{K,p,c}\cdot\boldsymbol{\mu}_{p,c}$
    \State $\mathcal{L}_{lb}\gets\operatorname{LoadBalanceLoss}(\boldsymbol{\pi})$
    \State $\mathcal{L}_{dppo}\gets\operatorname{PairPreferenceLoss}(\boldsymbol{\pi},y)$
    \If{training}
      \ForAll{$(p,c)\in\Omega_K$}
        \State $\mathbf{M}_{p,c}\gets(1-\eta)\mathbf{M}_{p,c}+\eta\cdot\operatorname{detach}(\mathbf{q}_{agg})$
      \EndFor
      \State $\tau\gets\operatorname{clamp}(\tau-\gamma\nabla_{\tau}\mathcal{L}_{total},\tau_{min},\tau_{max})$
    \EndIf
    \State \Return $\mathbf{r},\mathbf{c}^{pop},\mathcal{L}_{lb},\mathcal{L}_{dppo}$
  \end{algorithmic}
\end{algorithm}

\subsection{Training Setup}
Table~\ref{tab:s1_hparams} reports all key hyperparameters for SMPD-video, Informs, and MicroLens.

\begin{table}[t]
  \centering
  \scriptsize
  \setlength{\tabcolsep}{2pt}
  \renewcommand{\arraystretch}{0.98}
  \caption{Dataset-specific hyperparameter configuration used in the final runs.}
  \label{tab:s1_hparams}
  \begin{tabularx}{\columnwidth}{>{\raggedright\arraybackslash}p{0.44\columnwidth}>{\centering\arraybackslash}X>{\centering\arraybackslash}X>{\centering\arraybackslash}X}
    \toprule
    Hyperparameter & \makecell{SMPD-\\video} & Informs & \makecell{Micro-\\Lens} \\
    \midrule
    Optimizer & Adam & Adam & Adam \\
    Base learning rate & $2.0\times10^{-4}$ & $1.0\times10^{-5}$ & $3.5\times10^{-4}$ \\
    Encoder learning rate for CLIP/UAE & $8.0\times10^{-6}$ & $4.0\times10^{-7}$ & $1.4\times10^{-5}$ \\
    Batch size & 32 & 32 & 32 \\
    Epochs & 350 & 350 & 350 \\
    Weight decay & $1\times10^{-5}$ & $1\times10^{-5}$ & $1\times10^{-5}$ \\
    Dropout rate & 0.1 & 0.1 & 0.1 \\
    Sampled visual frames $T$ & 100 & 100 & 31 \\
    User feature dimension $d_{user}$ & 9 & 6 & 6 \\
    Text feature dimension $d_{text}$ & 1024 & 768 & 1024 \\
    Top-$K$ routing & 5 & 5 & 5 \\
    Popularity partitions $P$ & 6 & 6 & 6 \\
    Topic clusters per partition $C$ & 32 & 32 & 32 \\
    Initial routing temperature $\tau$ & 1.0 & 1.0 & 1.0 \\
    SSM inner hidden dimension $d_h$ & 1536 & 1536 & 1536 \\
    $\Delta_{min}/\Delta_{max}$ & $0.01 / 1.0$ & $0.01 / 1.0$ & $0.001 / 0.1$ \\
    Base window size & 20 & 20 & 10 \\
    DPPO coefficient $\lambda_{dppo}$ & 0.10 & 0.05 & 0.10 \\
    DPPO beta $\beta_{dppo}$ & 0.5 & 0.3 & 0.5 \\
    Load-balance coefficients $(\gamma_{lb},\beta_{lb})$ & $(0.01,0.01)$ & $(0.01,0.01)$ & $(0.01,0.01)$ \\
    Cross-attention layers $L_{ca}$ & 2 & 2 & 2 \\
    Early-stop patience & 10 & 5 & 6 \\
    \bottomrule
  \end{tabularx}
\end{table}

For reproducibility, we use a fixed runtime stack comprising Python 3.10.19 compiled with GCC 11.2.0, PyTorch 2.10.0+cu128, CUDA 12.8, cuDNN 9.1 build 91002, and Tesla V100S-PCIE-32GB hardware with 31.7\,GB memory and SM 7.0. Core dependencies include open\_clip\_torch 3.2.0, sentence-transformers 5.3.0, transformers 5.3.0, and numpy 2.2.6. Automatic mixed precision with torch.\allowbreak amp.\allowbreak autocast is enabled, and bfloat16 is supported. Final metrics are averaged across three seeds, with mean values reported in the main paper and per-seed logs retained for auditing. In addition, the train/validation/test splits strictly follow each benchmark's official protocol and are not reshuffled across methods.

\FloatBarrier

\section{Efficiency Analysis}

\subsection{Efficiency Advantage}
The 5.43$\times$ latency advantage over MMRA in Table~\ref{tab:s2_latency} is primarily rooted in the retrieval operator itself. At the algorithmic level, \method{} performs routing on a compact topology-aware prototype grid and aggregates only a sparse subset of memory slots, so the dominant retrieval cost is governed by a fixed structured memory. MMRA, by contrast, relies on dense interaction with a flat memory bank and therefore must process the full prototype set before effective retrieval signals can be formed. This difference shortens the critical inference path of \method{} and suppresses redundant global matching.

This algorithmic distinction further translates into a systems-level advantage. Because the memory tensor in \method{} remains compact and repeatedly reusable across queries, retrieval remains cache efficient and requires relatively limited intermediate feature movement. Attention over a flat memory bank exhibits weaker data reuse and heavier key/value traffic, making runtime more sensitive to memory movement than to arithmetic alone. This interpretation is also consistent with the stability results in Table~\ref{tab:s2_latency}: the lower coefficient of variation of \method{} indicates a more regular execution pattern, while MMRA is more vulnerable to memory-traffic fluctuations.

Overall, the latency and throughput gains in Table~\ref{tab:s2_latency} should be understood as a direct consequence of topology-aware sparse routing rather than as an isolated implementation artifact. The structured memory design reduces both redundant interaction and memory-access pressure, which jointly explains why \method{} is faster while remaining more stable.

\begin{table}[t]
  \centering
  \footnotesize
  \setlength{\tabcolsep}{3pt}
  \caption{Core efficiency comparison with batch size 16.}
  \label{tab:s2_latency}
  \begin{tabularx}{\columnwidth}{>{\raggedright\arraybackslash}p{0.34\columnwidth}>{\centering\arraybackslash}p{0.18\columnwidth}>{\centering\arraybackslash}p{0.18\columnwidth}>{\centering\arraybackslash}p{0.18\columnwidth}}
    \toprule
    Metric & \method{} & MMRA & ICPF \\
    \midrule
    Parameters & 1.28 M & 11.03 M & 4.91 M \\
    Mean latency & \textbf{7.28 ms} & 39.55 ms & 19.16 ms \\
    Std & \textbf{0.020 ms} & 0.538 ms & 0.105 ms \\
    P99 latency & \textbf{7.33 ms} & 40.58 ms & 19.59 ms \\
    Per-video latency & \textbf{0.455 ms} & 2.472 ms & 1.198 ms \\
    Throughput & \textbf{2198 FPS} & 405 FPS & 835 FPS \\
    Speedup vs. \method{} & 1.00$\times$ & 5.43$\times$ slower & 2.63$\times$ slower \\
    Coefficient of variation & \textbf{0.27\%} & 1.36\% & 0.55\% \\
    \bottomrule
  \end{tabularx}
\end{table}

\subsection{Scalability}
Table~\ref{tab:s3_batch_scaling} reports the scaling trend of \method{} as batch size increases.

\begin{table}[t]
  \centering
  \footnotesize
  \setlength{\tabcolsep}{3pt}
  \caption{Batch-size scalability of \method{}.}
  \label{tab:s3_batch_scaling}
  \begin{tabularx}{\columnwidth}{>{\centering\arraybackslash}p{0.10\columnwidth}>{\centering\arraybackslash}p{0.17\columnwidth}>{\centering\arraybackslash}p{0.20\columnwidth}>{\centering\arraybackslash}p{0.23\columnwidth}>{\centering\arraybackslash}p{0.18\columnwidth}}
    \toprule
    BS & Mean (ms) & Per-video (ms) & Throughput (FPS) & VRAM (GB) \\
    \midrule
    1 & 1.892 & 1.892 & 528.5 & 0.019 \\
    4 & 2.135 & 0.534 & 1872.9 & 0.022 \\
    8 & 3.242 & 0.405 & 2468.6 & 0.025 \\
    16 & 7.281 & 0.455 & 2197.6 & 0.032 \\
    32 & 14.563 & 0.455 & 2197.6 & 0.046 \\
    \bottomrule
  \end{tabularx}
\end{table}

Table~\ref{tab:s3_batch_scaling} further shows that this advantage is preserved as batch size increases. Once the fixed launch overhead is amortized, per-video latency remains nearly unchanged from medium batch sizes onward. This behavior is consistent with the design of \method{}: the routing stage reuses the same compact structured memory across samples, and its dominant cost is governed by a fixed prototype grid rather than by an expanding flat memory bank.

This property is important for practical deployment. Methods based on dense retrieval over flat memory banks typically face a sharper latency-throughput trade-off at larger batch sizes, because each additional sample still requires global interaction with a large memory set. In contrast, \method{} remains in a more favorable operating regime, sustaining high throughput without pronounced inflation in per-sample latency. Accordingly, the scalability trend in Table~\ref{tab:s3_batch_scaling} is fully consistent with the mechanism behind the latency advantage reported in Table~\ref{tab:s2_latency}.

\FloatBarrier

\section{Spatial Memory Analysis}

\subsection{Load-Balance Verification}
\begin{figure*}[t]
  \centering
  \includegraphics[width=0.98\textwidth]{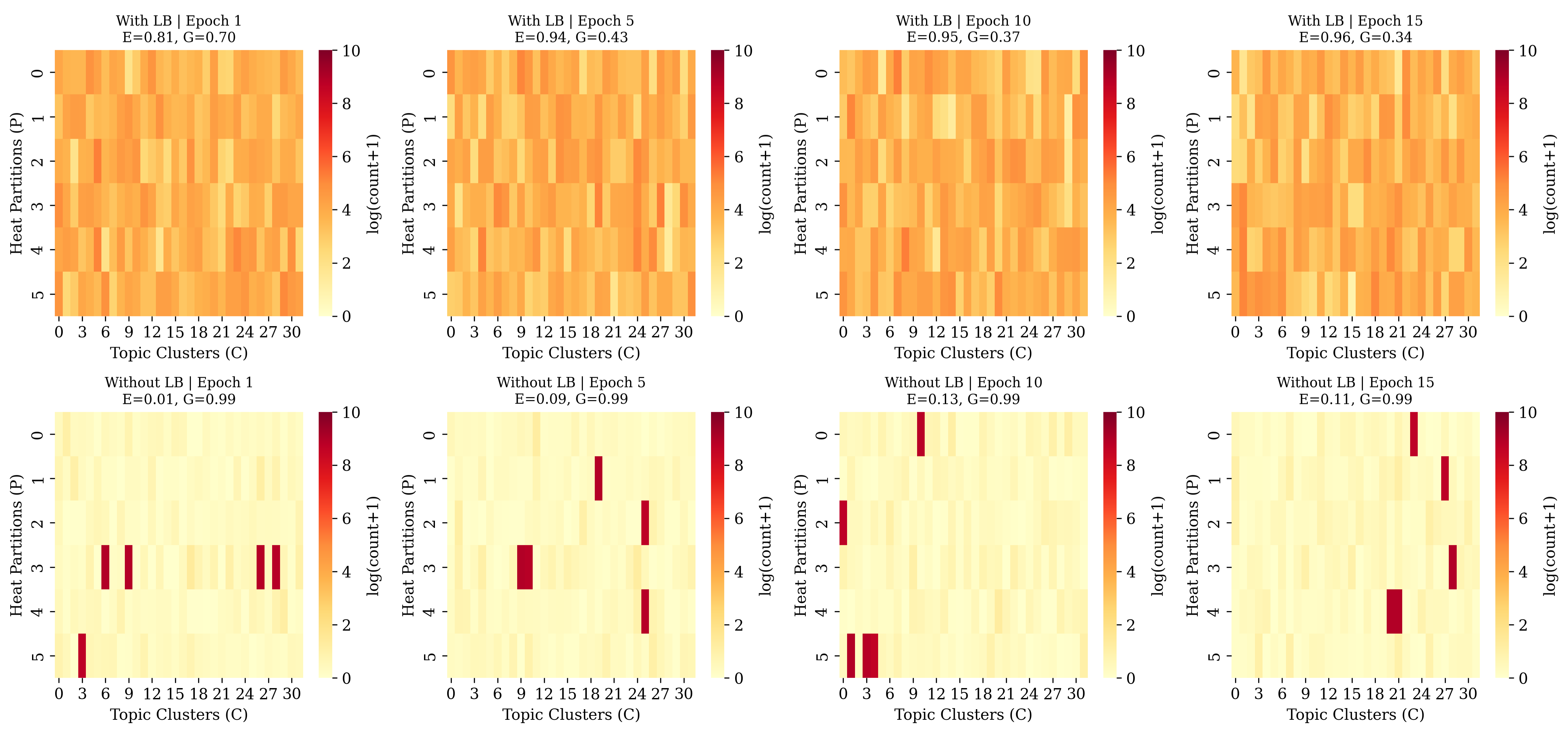}
  \caption{Slot activation heatmaps under topology-aware memory routing. Uniformly distributed activation over epochs indicates that the Load Balance Block prevents early-slot collapse.}
  \Description{Top row: slot-activation heatmaps with load balancing at Epochs 1, 5, 10, and 15. Bottom row: corresponding heatmaps without load balancing. Balanced routing produces diffuse activation across partitions and topic clusters, while removing load balancing leads to a small number of highly activated slots and severe collapse.}
  \label{fig:s1_slot_heatmap}
\end{figure*}

The role of the Load Balance Block in preventing slot collapse is first visible in the activation geometry of the memory bank. In Fig.~\ref{fig:s1_slot_heatmap}, when load balancing is enabled, slot activations remain broadly distributed over the $P\times C$ prototype grid from the earliest epochs onward, and no single partition-cluster pair dominates for long. The entropy annotated in the heatmaps rises from 0.81 at Epoch 1 to 0.96 at Epoch 15, while the corresponding Gini coefficient drops from 0.70 to 0.34, indicating that routing becomes progressively more uniform as training stabilizes.

By contrast, removing load balancing produces immediate concentration on a few highly activated slots. The lower row of Fig.~\ref{fig:s1_slot_heatmap} shows that the activation map without the Load Balance Block is sparse and highly concentrated across all observed epochs, with entropy remaining extremely low at 0.01--0.13 and the Gini coefficient fixed at 0.99. This pattern suggests that slot collapse is not merely a late-training artifact but an early optimization pathology in which a small subset of slots monopolizes retrieval traffic before the rest of the memory bank can specialize.

The same conclusion is supported quantitatively. Relative to the variant without load balancing, the full model more than doubles entropy from 0.41 to 0.87, reduces the Top-5 activation share from 0.78 to 0.29, and lowers the Gini coefficient from 0.63 to 0.21; moreover, collapse appears at Epoch 12 without load balancing, whereas it is absent in the full model. Overall, these statistics and Fig.~\ref{fig:s1_slot_heatmap} show that the Load Balance Block is not merely a regularizer but a structural condition for reliable topology-aware routing: it preserves memory coverage, avoids prototype starvation, and provides a stable spatial context for downstream popularity ranking.

\subsection{DPPO Convergence}
\begin{figure}[t]
  \centering
  \includegraphics[width=0.98\linewidth]{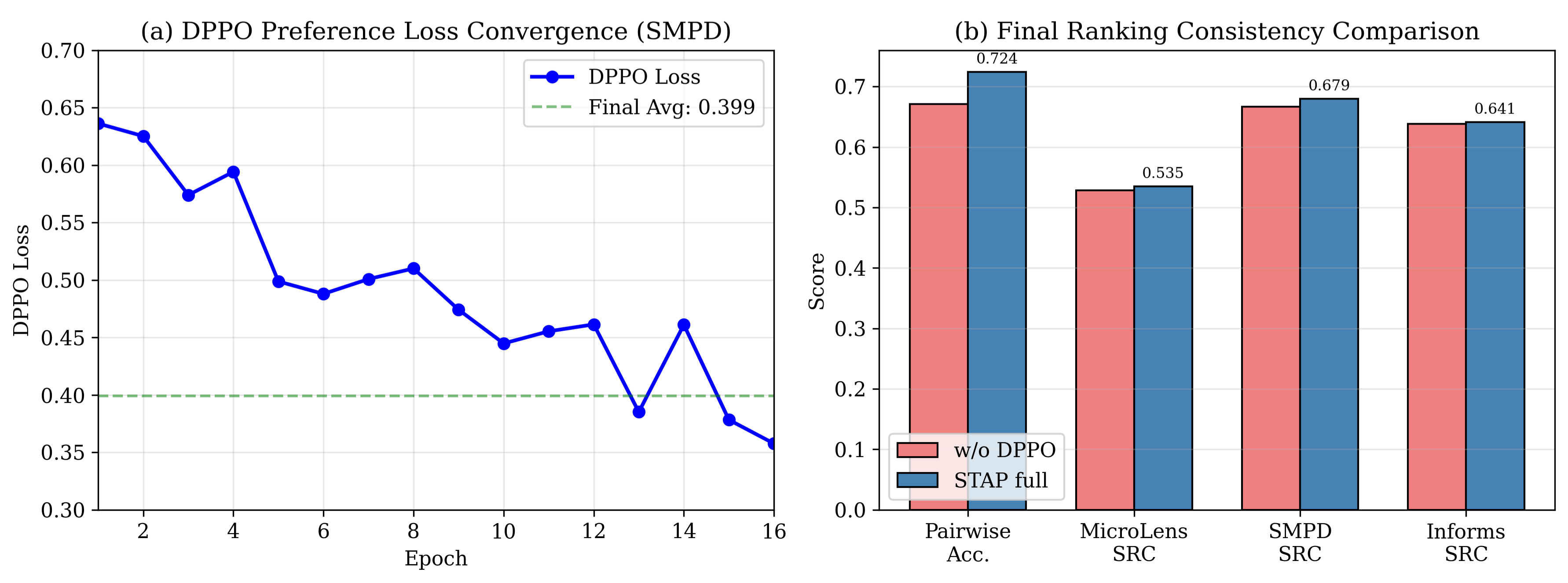}
  \caption{DPPO convergence behavior. The preference-alignment term converges smoothly and leads to more consistent final rankings.}
  \Description{Two-panel DPPO analysis. The left panel shows a steadily decreasing preference loss on SMPD-video over training epochs. The right panel compares final pairwise accuracy and SRC values on MicroLens, SMPD-video, and Informs for the variant without DPPO and the full model, showing consistent gains with DPPO.}
  \label{fig:s2_dppo_curve}
\end{figure}

The effect of DPPO can be understood from both optimization dynamics and final ranking quality. In Fig.~\ref{fig:s2_dppo_curve}(a), the preference loss on SMPD-video decreases from about 0.64 at the beginning of training to about 0.36 by Epoch 16, with only mild local oscillations and a late-stage average of 0.399. This trajectory indicates that the pairwise preference objective converges smoothly without conflicting with the main regression objective, and that the model gradually learns a more consistent ordering over popularity pairs.

This optimization behavior translates into a measurable ranking benefit across datasets. As shown in Fig.~\ref{fig:s2_dppo_curve}(b), adding DPPO increases pairwise accuracy from 0.671 to 0.724, while SRC improves from 0.5282 to 0.5346 on MicroLens, from 0.6666 to 0.6794 on SMPD-video, and from 0.2685 to 0.2813 on Informs. The corresponding relative gains are 7.9\%, 1.2\%, 1.9\%, and 4.8\%, respectively, showing that the preference signal not only refines local pair ordering but also improves global ranking coherence under different content distributions.

Overall, the convergence curve and the cross-dataset ranking gains indicate that DPPO serves as a stabilizing alignment objective for the spatial memory bank. By encouraging retrieved prototypes to respect pairwise popularity preference, DPPO suppresses noisy routing outcomes and makes the retrieved spatial context more discriminative, which is consistent with the main-text explanation of stronger spatial robustness under cold-start retrieval.

\subsection{Grid Search over P and C}
\begin{figure*}[t]
  \centering
  \includegraphics[width=0.98\textwidth]{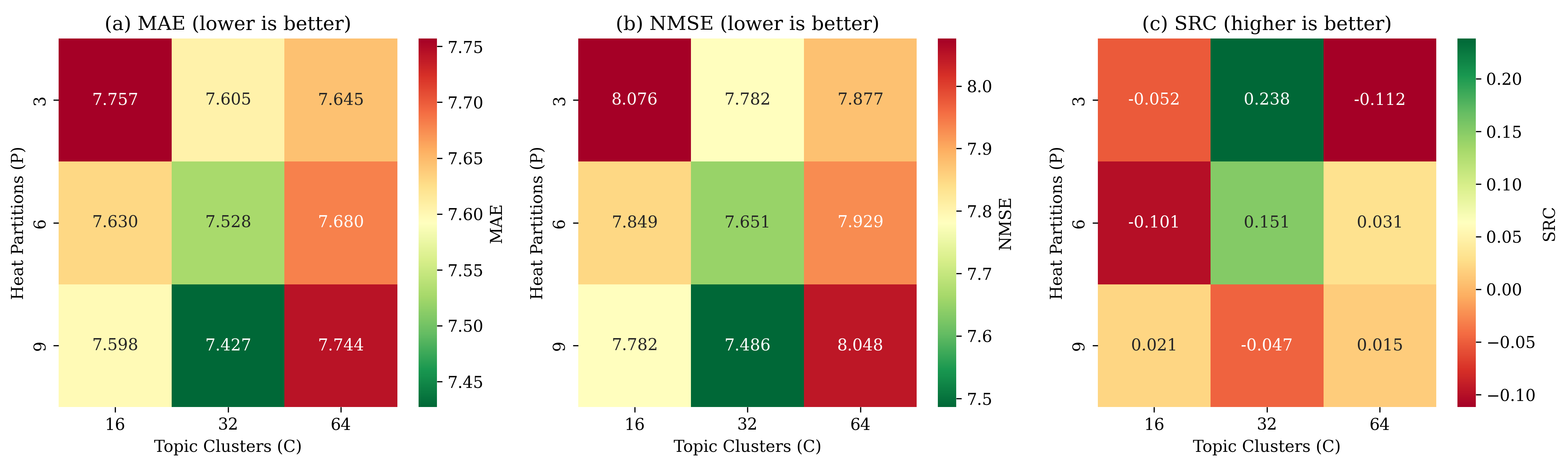}
  \caption{Grid search heatmap over $P\times C$ configurations. An intermediate memory-bank scale yields the most favorable overall trade-off, and the final model uses $P{=}6$ and $C{=}32$, corresponding to 192 slots.}
  \Description{Three heatmaps compare MAE, nMSE, and SRC across combinations of heat partitions P and topic clusters C. The configuration with P equal to 6 and C equal to 32 provides the most favorable balance across metrics.}
  \label{fig:s3_pc_grid_heatmap}
\end{figure*}

We further examine how memory-bank scale affects retrieval quality by jointly varying the numbers of heat partitions $P$ and topic clusters $C$. Fig.~\ref{fig:s3_pc_grid_heatmap} visualizes MAE, nMSE, and SRC over the $P\times C$ grid and reveals a clearly non-monotonic trend: enlarging the bank helps only up to a certain point, after which the benefit of additional prototypes is offset by sparser updates and noisier routing. The strongest overall balance is achieved at $P{=}6$ and $C{=}32$, corresponding to 192 slots, with MAE 7.528, nMSE 7.651, and SRC 0.151. This result indicates that bank scale must match the effective routing density of the model: the memory should be large enough to preserve cross-video prototype diversity, but not so large that each slot receives insufficient support during training.

This result can be understood through the same lens as the previous two subsections. The Load Balance Block prevents early-slot collapse and keeps routing traffic distributed across the bank, while DPPO aligns the retrieved prototypes with pairwise popularity preference and suppresses noisy retrieval outcomes. However, these two mechanisms do not eliminate the need for an appropriate bank size. If the grid is too small, even balanced routing still compresses heterogeneous samples into overly coarse slots; if the grid is too large, even well-regularized routing becomes fragmented and each slot is updated by too few compatible samples to specialize reliably.

Table~\ref{tab:s4_pc_grid} further clarifies the behavior of undersized memory banks. Among the smallest configurations, $3\times16$, corresponding to 48 slots, performs worst overall, with MAE 7.757, nMSE 8.076, and SRC $-0.052$, showing that the bank cannot cover sufficiently diverse popularity-topic structure. Expanding to $3\times32$ improves ranking discrimination and yields the highest SRC of 0.238, but its MAE and nMSE remain worse than those of the selected setting. This result indicates that a small bank may sharpen relative ordering for part of the data, yet still lacks sufficient prototype diversity to provide stable global support for regression and retrieval.

Larger grids reveal the complementary limitation. Although $9\times32$ attains the lowest MAE and nMSE, namely 7.427 and 7.486, its SRC drops to $-0.047$, and both $6\times64$ and $9\times64$ degrade across all three metrics. This pattern indicates that increasing the number of slots does not monotonically improve memory routing. Once the grid becomes overly fine-grained, the sample support per slot decreases, prototype specialization becomes less reliable, and routing increasingly absorbs idiosyncratic memory noise rather than robust cross-video structure. Overall, Fig.~\ref{fig:s3_pc_grid_heatmap} and Table~\ref{tab:s4_pc_grid} indicate that an intermediate grid scale is the regime in which topology-aware memory is most effective. The selected $6\times32$ bank is large enough to cover meaningful popularity-topic structure, yet compact enough to keep each slot sufficiently trained and consistently routed. This finding is consistent with the conclusions of the previous two subsections: after load balancing prevents slot collapse and DPPO improves preference alignment, an intermediate bank size provides the most favorable balance among memory diversity, routing stability, and ranking coherence.

\begin{table}[t]
  \centering
  \scriptsize
  \setlength{\tabcolsep}{2pt}
  \renewcommand{\arraystretch}{0.98}
  \caption{One-column summary of representative $P\times C$ configurations. An intermediate bank size provides the most favorable overall trade-off.}
  \label{tab:s4_pc_grid}
  \begin{tabularx}{\columnwidth}{>{\centering\arraybackslash}p{0.12\columnwidth}>{\centering\arraybackslash}p{0.11\columnwidth}>{\centering\arraybackslash}p{0.12\columnwidth}>{\centering\arraybackslash}p{0.11\columnwidth}>{\centering\arraybackslash}p{0.10\columnwidth}>{\raggedright\arraybackslash}X}
    \toprule
    $P\times C$ & MAE$\downarrow$ & nMSE$\downarrow$ & SRC$\uparrow$ & Slots & Comment \\
    \midrule
    $3\times16$ & 7.757 & 8.076 & -0.052 & 48 & Insufficient capacity \\
    $3\times32$ & 7.605 & 7.782 & 0.238 & 96 & Highest SRC among small grids \\
    $3\times64$ & 7.645 & 7.877 & -0.112 & 192 & Sparse slot updates \\
    $6\times16$ & 7.630 & 7.849 & -0.101 & 96 & Limited ranking stability \\
    $6\times32$ & 7.528 & 7.651 & 0.151 & 192 & \textbf{Selected configuration} \\
    $6\times64$ & 7.680 & 7.929 & 0.031 & 384 & Higher capacity, lower accuracy \\
    $9\times16$ & 7.598 & 7.782 & 0.021 & 144 & Moderate capacity, limited SRC \\
    $9\times32$ & 7.427 & 7.486 & -0.047 & 288 & Lowest MAE/nMSE, unstable SRC \\
    $9\times64$ & 7.744 & 8.048 & 0.015 & 576 & Evidence of overfitting \\
    \bottomrule
  \end{tabularx}
\end{table}

\FloatBarrier

\section{Frame-Level Visualization}

\subsection{Case Studies}
\begin{figure*}[!b]
  \centering
  \includegraphics[width=0.98\textwidth]{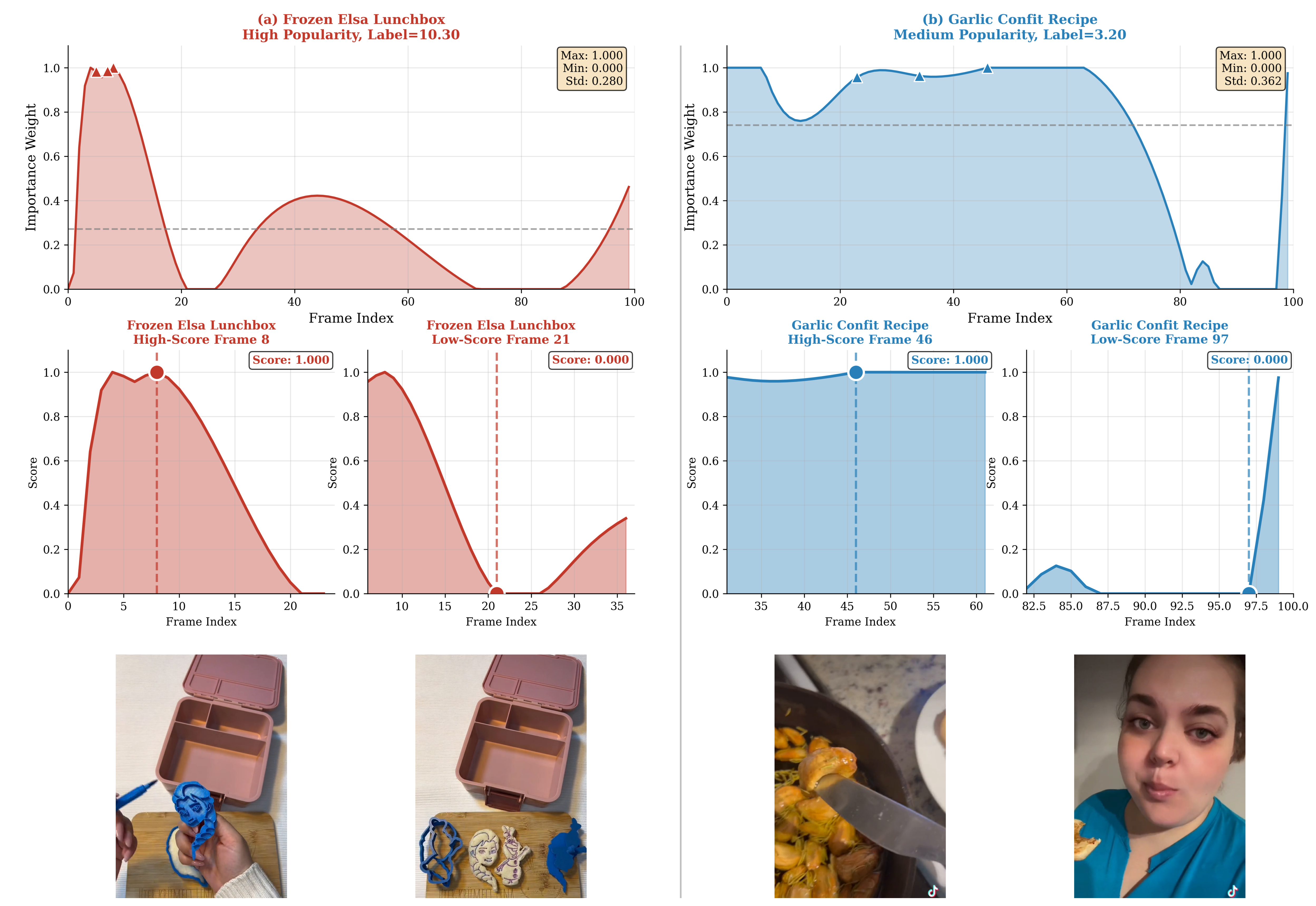}
  \caption{Combined case studies and peak-frame alignment for the frame-scoring module. The left column shows a high-popularity lunchbox video with a strong early peak and a low-score static frame; the right column shows a medium-popularity recipe video with sustained high importance over the cooking segment and a low-score frame from a late speaking segment. Across both cases, high-score frames correspond to semantically decisive moments, while low-score frames are relatively static, repetitive, or weakly relevant to the main content.}
  \Description{A combined figure with two video cases. For each case, the top panel shows the full frame-importance curve, the middle panels highlight one high-score frame and one low-score frame index, and the bottom panels show the corresponding video snapshots.}
  \label{fig:s3_case_alignment}
\end{figure*}

Figure~\ref{fig:s3_case_alignment} presents two representative examples: \textit{Frozen Elsa Lunchbox}, a high-popularity sample with label 10.30, and \textit{Garlic Confit Recipe}, a medium-popularity sample with label 3.20. Instead of assigning uniform temporal importance, \method{} produces structured weighting patterns that follow the semantic progression of each video. In the lunchbox example, the score rises sharply at the beginning and reaches its maximum at Frame 8, indicating that the model strongly emphasizes the initial attention-capturing segment and the most visually distinctive hand-object interaction. The curve then decreases rapidly and later exhibits a weaker mid-video elevation, suggesting that the model captures secondary content cues while assigning them lower importance than the opening highlight.

The recipe example exhibits a different temporal profile. Its importance curve remains near saturation throughout an extended central interval, indicating that the cooking process provides sustained semantically relevant evidence rather than a few isolated peaks. The score then declines sharply in the later segment and reaches a minimum at Frame 97, implying that the terminal frames contribute substantially less to popularity prediction than the central cooking sequence. This contrast between concentrated early emphasis and sustained mid-video emphasis indicates that the frame-scoring module adapts to different temporal structures rather than enforcing a fixed pattern across videos.

\subsection{Peak-Frame Alignment}
To provide direct visual evidence, Fig.~\ref{fig:s3_case_alignment} aligns local maxima and minima of the scoring curve with their corresponding frame screenshots. In the lunchbox example, the top-scoring frame captures active manipulation of the themed lunchbox elements, whereas the low-scoring frame is comparatively static and compositionally repetitive. In the recipe example, the top-scoring frame focuses on the core cooking content, while the low-scoring frame appears in the late speaking segment and is much less aligned with the dominant recipe semantics. These matched comparisons indicate that local peaks are not numerical artifacts; rather, they consistently correspond to frames with stronger action cues, clearer narrative function, or richer content information.

Overall, these case studies show that \method{} emphasizes visually and semantically decisive moments while suppressing temporally uninformative segments such as static transitions, repeated layouts, or weakly relevant ending shots. This frame-level evidence directly supports the temporal modeling claim: \method{} achieves stronger aggregate performance while remaining behaviorally interpretable through its temporal emphasis patterns.

\FloatBarrier